\documentclass[]{jfm}

\usepackage[dvipsnames]{xcolor}

\usepackage{graphicx}
\usepackage{newtxtext}
\usepackage{newtxmath}
\usepackage{natbib}
\usepackage{hyperref}
\hypersetup{
    colorlinks = true,
    urlcolor   = blue,
    citecolor  = black,
}

\newcommand{\RomanNumeralCaps}[1]
\linenumbers

\usepackage{graphicx}
\usepackage{epstopdf, epsfig}
\usepackage{amsmath}
\usepackage[utf8]{inputenc} 

\usepackage{subfigure}

\usepackage{pgfplots}
\usepackage{tikz}
\usetikzlibrary{backgrounds,pgfplots.groupplots,external}

\title{A revised gap-averaged Floquet analysis of Faraday waves in Hele-Shaw cells} 

\author{Alessandro Bongarzone\aff{1} \corresp{\email{alessandro.bongarzone@epfl.ch}},
  Baptiste Jouron,
  Francesco Viola\aff{2}
 \and François Gallaire\aff{1}
 }

\affiliation{\aff{1}Laboratory of Fluid Mechanics and Instabilities, École Polytechnique Fédérale de Lausanne, Lausanne, CH-1015, Switzerland
\aff{2}Gran Sasso Science Institute, Viale F. Crispi, 7, 67100 L'Aquila, Italy}

\begin{document}

\maketitle

\begin{abstract}

Existing theoretical analyses of Faraday waves in Hele-Shaw cells rely on the Darcy approximation and assume a parabolic flow profile in the narrow direction. However, Darcy's model is known to be inaccurate when convective or unsteady inertial effects are important. In this work, we propose a gap-averaged Floquet theory accounting for inertial effects induced by the unsteady terms in the Navier-Stokes equations, a scenario that corresponds to a pulsatile flow where the fluid motion reduces to a two-dimensional oscillating Poiseuille flow, similarly to the Womersley flow in arteries. When gap-averaging the linearized Navier-Stokes equation, this results in a modified damping coefficient, which is a function of the ratio between the Stokes boundary layer thickness and the cell's gap, and whose complex value depends on the frequency of the wave response specific to each unstable parametric region. We first revisit the standard case of horizontally infinite rectangular Hele-Shaw cells by also accounting for a dynamic contact angle model. A comparison with existing experiments shows the predictive improvement brought by the present theory and points out how the standard gap-averaged model often underestimates the Faraday threshold. The analysis is then extended to the less conventional case of thin annuli. A series of dedicated experiments for this configuration highlights how Darcy's thin-gap approximation overlooks a frequency detuning that is essential to correctly predict the locations of the Faraday tongues in the frequency-amplitude parameter plane. These findings are well rationalized and captured by the present model.  

\end{abstract}

\section{Introduction}\label{sec:Sec0}

\indent Recent Hele-Shaw cell experiments have enriched the knowledge of Faraday waves \citep{Faraday1831}. Researchers have uncovered a new type of highly localized standing waves, referred to as oscillons, that are both steep and solitary-like in nature \citep{rajchenbach2011new}. These findings have spurred further experimentations with Hele-Shaw cells filled with one or more liquid layers, using a variety of fluids, ranging from silicone oil, and water-ethanol mixtures to pure ethanol \citep{li2018effect}. Through these experiments, new combined patterns produced by triadic interactions of oscillons were discovered by \cite{li2014observations}. Additionally, another new family of waves was observed in a cell filled solely with pure ethanol and at extremely shallow liquid depths \citep{li2015observation,li2016pattern}.\\
\indent All these findings contribute to the understanding of the wave behaviour in Hele-Shaw configurations and call for a reliable stability theory that can explain and predict the instability onset for the emergence of initial wave patterns.\\
\indent Notwithstanding two-dimensional direct numerical simulations \citep{perinet2016hysteretic,ubal2003numerical} have been able to qualitatively replicate standing wave patterns reminiscent of those observed in experiments \citep{li2014observations}, these simulations overlook the impact of wall attenuation, hence resulting in a simplified model that cannot accurately predict the instability regions \citep{Benjamin54,kumar1994parametric} and is therefore not suitable for modelling Hele-Shaw flows. On the other hand, attempting to conduct three-dimensional simulations of fluid motions in a Hele-Shaw cell poses a major challenge due to the high computational cost associated with the narrow dimension of the cell, which requires a smaller grid cell size to capture the shear dissipation accurately. Consequently, the cost of performing such simulations increases rapidly.\\
\indent In order to tackle the challenges associated with resolving fluid dynamics within such systems, researchers have utilized Darcy's law as an approach to treating the confined fluid between two vertical walls. This approximation, also used in the context of porous medium, considers the fluid to be flowing through a porous medium, resulting in a steady parabolic flow in the short dimension. When gap-averaging the linearized Navier-Stokes equation, this approximation translates into a damping coefficient $\sigma$ that scales as $12\nu/b^2$, with $\nu$ the fluid kinematic viscosity and $b$ the cell's gap-size, which represents the boundary layer dissipation at the lateral walls. However, Darcy's model is known to be inaccurate when convective and unsteady inertial effects are not negligible, such as in waves \citep{kalogirou2016variational}. It is challenging to reintroduce convective terms consistently into the gap-averaged Hele-Shaw equations from a mathematical standpoint \citep{ruyer2001inertial,plouraboue2002kelvin,luchini2010consistent}.\\
\indent In their research, \cite{li2019stability} applied the Kelvin-Helmholtz-Darcy theory proposed by \cite{gondret1997shear} to reintroduce advection and derive the nonlinear gap-averaged Navier-Stokes equations. These equations were then implemented in the open-source code \textit{Gerris} developed by \citep{popinet2003gerris,popinet2009accurate} to simulate Faraday waves in a Hele-Shaw cell. Although this gap-averaged model was compared to several experiments and demonstrated fairly good agreement, it should be noted that the surface tension term remains two-dimensional, as the out-of-plane interface shape is not directly taken into account. This simplified treatment neglects the contact line dynamics and may lead to miscalculations in certain situations. Advances in this direction were made by \cite{li2018faraday}, who found that the out-of-plane capillary forces associated with the meniscus curvature across the thin-gap direction should be retained in order to improve the description of the wave dynamics, as experimental evidence suggests. By employing a more sophisticated model, coming from molecular kinetics theory \citep{blake1993dynamic,hamraoui2000can,blake2006physics} and similar to the macroscopic one introduced by \cite{Hocking87}, to include the capillary contact line motion arising from the small scale of the gap-size between the two walls of a Hele-Shaw cell, they derived a novel dispersion relation, which indeed better predicts the observed instability onset.\\
\indent However, discrepancies in the instability thresholds were still found. This mismatch was tentatively attributed to factors that are not accounted for in the gap-averaged model, such as the extra dissipation on the lateral walls in the elongated direction. Of course, a lab-scale experiment using a rectangular cell cannot entirely replace an infinite-length model, but if the container is sufficiently long, then this extra dissipation should be negligible. Other candidates were identified in the phenomenological contact line model or free surface contaminations.\\
\indent If these factors can certainly be sources of discrepancies, we believe that a pure hydrodynamic effect could be at the origin of the discordance between theory and experiments in the first place.\\
\indent Despite the use of the Darcy approximation is well-assessed in the literature, the choice of a steady Poiseuille flow profile as an ansatz to build the gap-averaged model appears in fundamental contrast with the unsteady nature of oscillatory Hele-Shaw flows, such as Faraday waves. At low enough oscillation frequencies or for sufficiently viscous fluids, the thickness of the oscillating Stokes boundary layer becomes comparable to the cell gap: the Stokes layers over the lateral solid faces of the cell merge and eventually invade the entire fluid bulk. In such scenarios, the Poiseuille profile gives an adequate flow description, but this pre-requisite is rarely met in the above-cited experimental campaigns. It appears, thus, very natural to ask oneself whether a more appropriate description of the oscillating boundary layer impacts the prediction of stability boundaries. This study is precisely devoted to answering this question by proposing a revised gap-averaged Floquet analysis, based on the classical Womersley-like solution for the pulsating flow in a channel \citep{womersley1955method,san2012improved}.\\
\indent Following the approach taken by \cite{viola2017sloshing}, we examine the impact of inertial effects on the instability threshold of Faraday waves in Hele-Shaw cells, with a focus on the unsteady term of the Navier-Stokes equations. This scenario corresponds to a pulsatile flow where the fluid’s motion reduces to a two-dimensional oscillating channel flow, which seems better suited than the steady Poiseuille profile to investigate the stability properties of the system. When gap-averaging the linearized Navier-Stokes equation, this results in a modified damping coefficient becoming a function of the ratio between the Stokes boundary layer thickness and the cell’s gap, and whose complex value will depend on the frequency of the wave response specific to each unstable parametric region.\\
\indent First, we consider the case of horizontally infinite rectangular Hele-Shaw cells by also accounting for the same dynamic contact angle model employed by \cite{li2019stability}, so as to quantify the predictive improvement brought by the present theory. A \textit{vis}-à-\text{vis} comparison with experiments by \cite{li2019stability} points out how the standard Darcy model often underestimates the Faraday threshold, whereas the present theory can explain and close the gap with these experiments.\\
\indent The analysis is then extended to the case of thin annuli. This less common configuration has been already used to investigate oscillatory phase modulation of parametrically forced surface waves \citep{douady1989oscillatory} and drift instability of cellular patterns \citep{fauve1991drift}. For our interest, an annular cell is convenient as it naturally filters out the extra dissipation that could take place on the lateral boundary layer in the elongated direction, hence allowing us to reduce the sources of extra uncontrolled dissipation and perform a cleaner comparison with experiments. Our homemade experiments for this configuration highlight how Darcy's theory overlooks a frequency detuning that is essential to correctly predict the locations of the Faraday's tongues in the frequency spectrum. These findings are well rationalized and captured by the present model.\\
\indent The paper is organized as follows. In \S\ref{sec:C8_Sec1} we revisit the classical case of horizontally infinite rectangular Hele-Shaw cells. The present model is compared with theoretical predictions from the standard Darcy theory and with existing experiments. The case of thin annuli is then considered. The model for the latter unusual configuration is formulated in \S\ref{sec:C8_Sec2} and compared with homemade experiments in \S\ref{sec:C8_EXP}. Conclusions are outlined in \S\ref{sec:C8_CONC}.

\section{Horizontally infinite Hele-Shaw cells}\label{sec:C8_Sec1}

\begin{figure}
\centering
\includegraphics[width=0.8\textwidth]{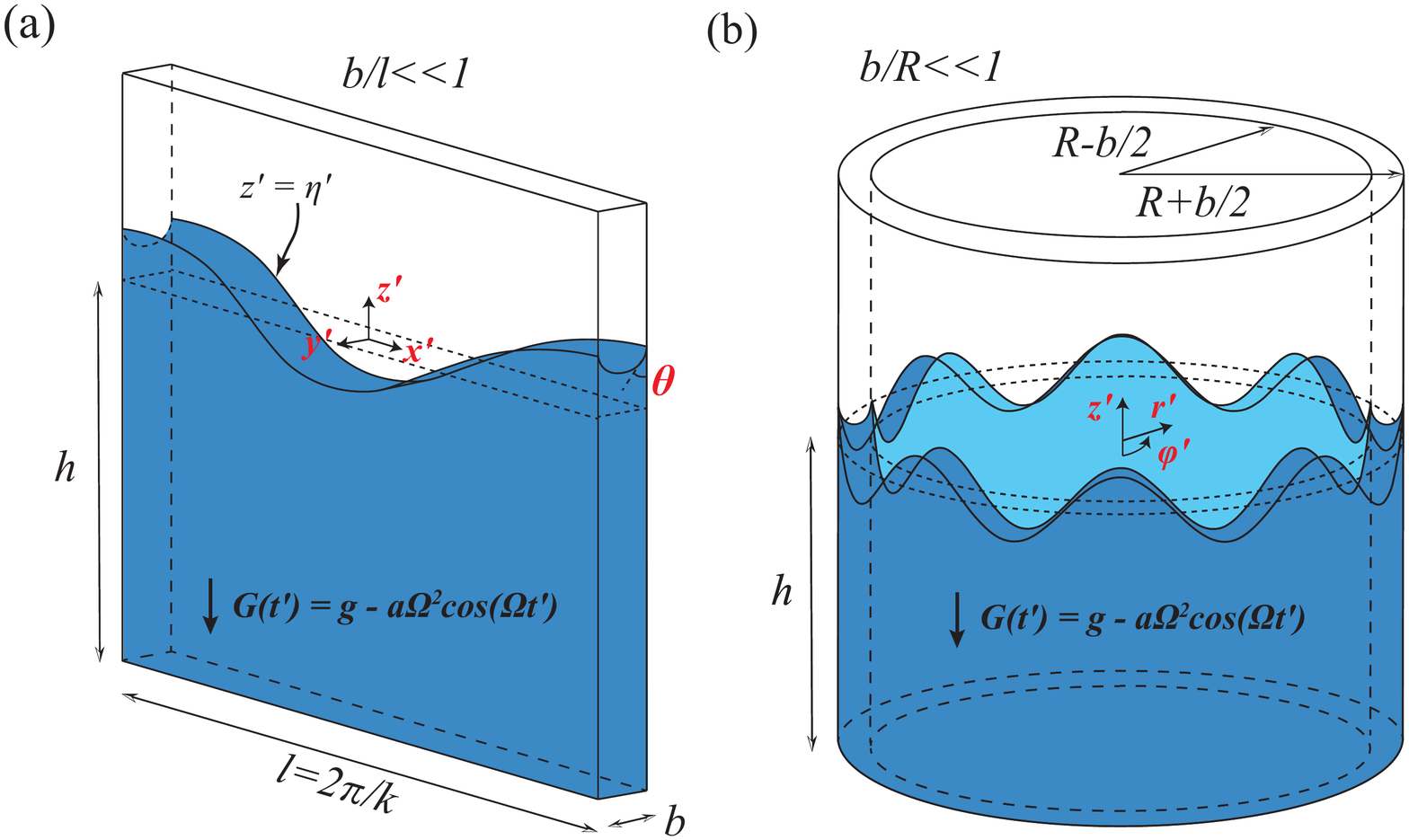}
\caption{(a) Sketch of Faraday waves in a rectangular Hele-Shaw cell of width $b$ and length $l$ filled to a depth $h$ with a liquid. Here $b$ denotes the gap size of the Hele-Shaw cell, $l$ is the wavelength of a certain wave, such that $b/l\ll 1$, and $\theta$ is the dynamic contact angle of the liquid on the lateral walls. The vessel undergoes a vertical sinusoidal oscillation of amplitude $a$ and angular frequency $\Omega$. The free surface elevation is denoted by $\eta'\left(x'\right)$. (b) Same as (a), but in an annular Hele-Shaw cell with internal and external radii, respectively, $R-b/2$ and $R+b/2$. Here, $b/R\ll 1$ and the free surface elevation is a function of the azimuthal coordinate $\varphi'$, i.e. $\eta'\left(\varphi'\right)$.}
\label{fig:C8_Sketch1} 
\end{figure}

\noindent Let us begin by considering the case of a horizontally infinite Hele-Shaw cell of width $b$ filled to a depth $h$ with an incompressible fluid of density $\rho$, dynamic viscosity $\mu$ (kinematic viscosity $\nu=\mu/\rho$) and liquid-air surface tension $\gamma$ (see also sketch in figure~\ref{fig:C8_Sketch1}(a)). The vessel undergoes a vertical sinusoidal oscillation of amplitude $a$ and angular frequency $\Omega$. In a frame of reference which moves with the oscillating container, the free liquid interface is flat and stationary for small forcing amplitudes, and the oscillation is equivalent to a temporally modulated gravitational acceleration, $G\left(t'\right)=g-a\Omega^2\cos{\Omega t'}$. The equation of motion for the fluid bulk are
\begin{equation}
    \label{eq:C8_NSfull}
    \rho\left(\frac{\partial \mathbf{U}'}{\partial t'}+\mathbf{U'}\cdot\nabla'\mathbf{U'}\right)=-\nabla' P' + \mu\nabla'^2\mathbf{U'}-\rho G\left(t\right)\mathbf{e}_z,\ \ \ \ \nabla'\cdot\mathbf{U'}=0.
\end{equation}
\noindent Linearizing about the rest state $\mathbf{U}'=\mathbf{0}$ and $P'\left(z',t'\right)=-\rho G\left(t'\right) z'$, the equations for the perturbation velocity, $\mathbf{u}'\left(x',y',z',t'\right)=\left\{u',v',w'\right\}^T$, and pressure, $p'\left(x',y',z',t'\right)$, fields, associated with a certain perturbation's wavelength $l~\sim k^{-1}$ ($k$, wavenumber), read
\begin{equation}
    \label{eq:C8_NSlin}
    \rho\frac{\partial\mathbf{u}'}{\partial t'}=-\nabla' p' + \mu\nabla^2\mathbf{u}',\ \ \ \ \nabla'\cdot\mathbf{u}'=0.
\end{equation}
\noindent Assuming that $bk\ll 1$, then the velocity along the narrow $y'$-dimension $v'\ll u',w'$ and, by employing the Hele-Shaw approximation as in, for instance, \cite{viola2017sloshing}, one can simplify the linearized Navier-Stokes equations as follows:
\begin{subequations}
\begin{equation}
    \label{eq:C8_NS_HS_cont}
    \frac{\partial u'}{\partial x'}  + \frac{\partial v'}{\partial y'} + \frac{\partial w'}{\partial z'} = 0,
\end{equation}
\begin{equation}
    \label{eq:C8_NS_HS_uvw}
    \rho\frac{\partial u'}{\partial t'} = -\frac{\partial p'}{\partial x'} + \mu \frac{\partial^2 u'}{\partial y'^2},\ \ \ \ \rho\frac{\partial w'}{\partial t'} = -\frac{\partial p'}{\partial z'} + \mu \frac{\partial^2 w'}{\partial y'^2},\ \ \ \ \frac{\partial p'}{\partial y'}=0.
\end{equation}
\end{subequations}
\noindent Equations~\eqref{eq:C8_NS_HS_cont}-\eqref{eq:C8_NS_HS_uvw} are made dimensionless using $k^{-1}$ for the directions $x$ and $z$, and $b$ for $y$. The forcing amplitude and frequency provide a scale $a\Omega$ for the in-plane $xz$-velocity components, whereas the continuity equation imposes the transverse component $v'$  to scale as $bk a\Omega\ll a\Omega$, due to the strong confinement in the $y$-direction ($bk\ll 1$). With these
choices, dimensionless spatial scales, velocity components and pressure write:
\begin{equation}
    \label{eq:C8_scaling_nondim}
    x=x'k,\ \ \ y=\frac{y'}{b},\ \ \ z=z' k,\ \ \ u=\frac{u'}{a\Omega^2},\ \ \ v=\frac{v'}{bk a\Omega},\ \ \ w=\frac{w'}{a\Omega},\ \ \ p=\frac{kp'}{\rho a\Omega^2}.
\end{equation}
\noindent The first two equations in~\eqref{eq:C8_NS_HS_uvw} in a non-dimensional form are
\begin{equation}
        \label{eq:C8_NS_HS_nodim_uv}
        \frac{\partial u}{\partial t}=-\frac{\partial p}{\partial x}+\frac{\delta_{St}^2}{2} \frac{\partial^2 u}{\partial y^2},\ \ \ \ \ \frac{\partial w}{\partial t}=-\frac{\partial p}{\partial z}+\frac{\delta_{St}^2}{2} \frac{\partial^2 w}{\partial y^2},
\end{equation}
\noindent where $\delta_{St}=\delta_{St}'/b$ and with $\delta_{St}'=\sqrt{2\nu/\Omega}$ denoting the thickness of the oscillating Stokes boundary layer. The ratio $\sqrt{2}/\delta_{St}$ is also commonly referred to as the Womersley number, $Wo=b\sqrt{\Omega/\nu}$ \citep{womersley1955method,san2012improved}.

\subsection{Floquet analysis of the gap-averaged equations}\label{subsec:C8_Sec1sub2}

\begin{figure}
\centering
\includegraphics[width=0.9\textwidth]{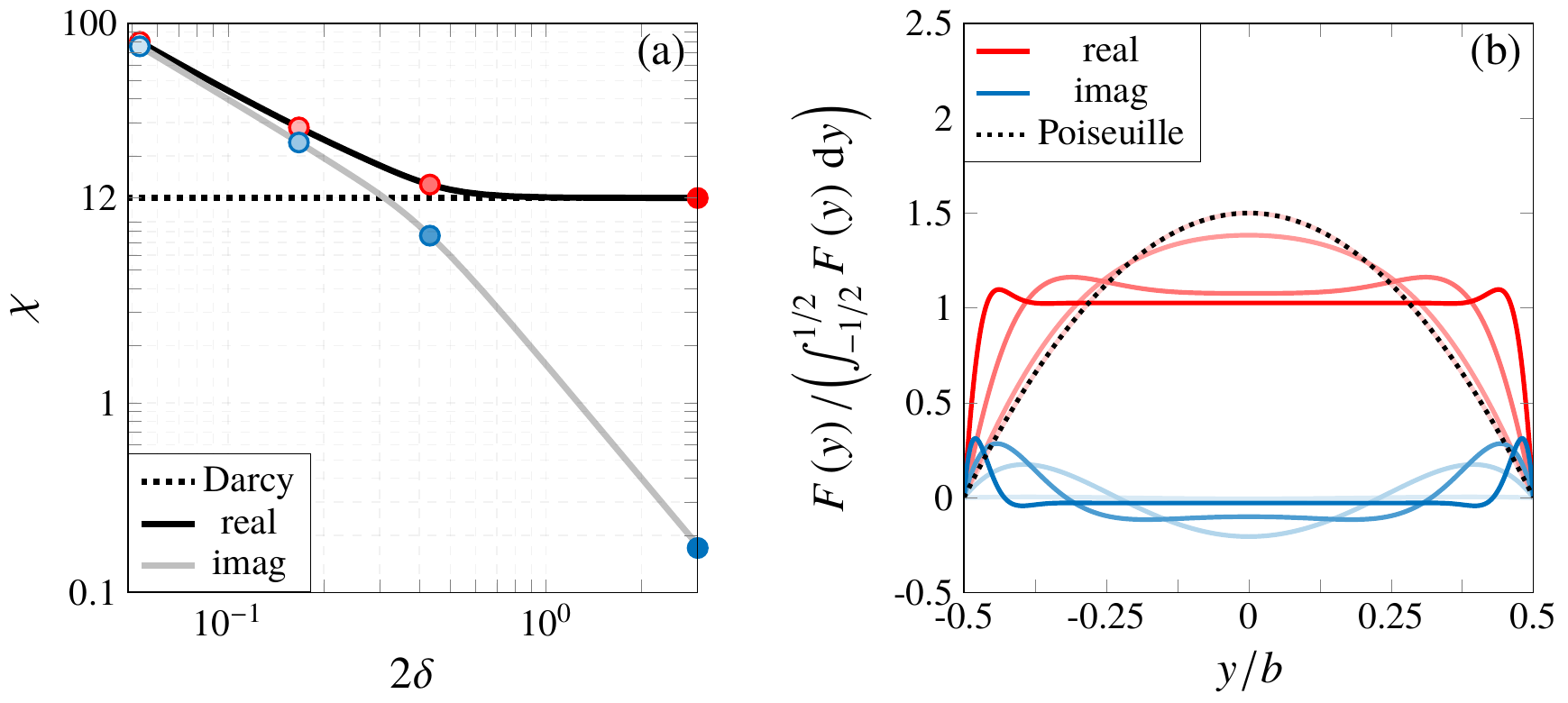}
\caption{(a) Real and imaginary parts of the complex auxiliary coefficient $\chi=\chi_r+\text{i}\chi_i$ versus twice the non-dimensional Stokes boundary layer thickness $\delta$. The horizontal black dotted line indicates the constant value $12$ given by the Darcy approximation. (b) Normalized profile $F\left(y\right)$ (Womersley profile) for different $\delta=b^{-1}\sqrt{2\nu/\xi\Omega}$, whose values are specified by the filled circles in (a) with matching colors. The Poiseuille profile is also reported for completeness. In drawing these figures we let the oscillation frequency of the wave, $\xi\Omega$, free to assume any value, but we recall that the parameter $\xi$ can only assume discrete values, and so do $\chi$ and $F\left(y\right)$.}
\label{fig:C8_profile_fz_chin} 
\end{figure}

\begin{figure}
\centering
\includegraphics[width=1.0\textwidth]{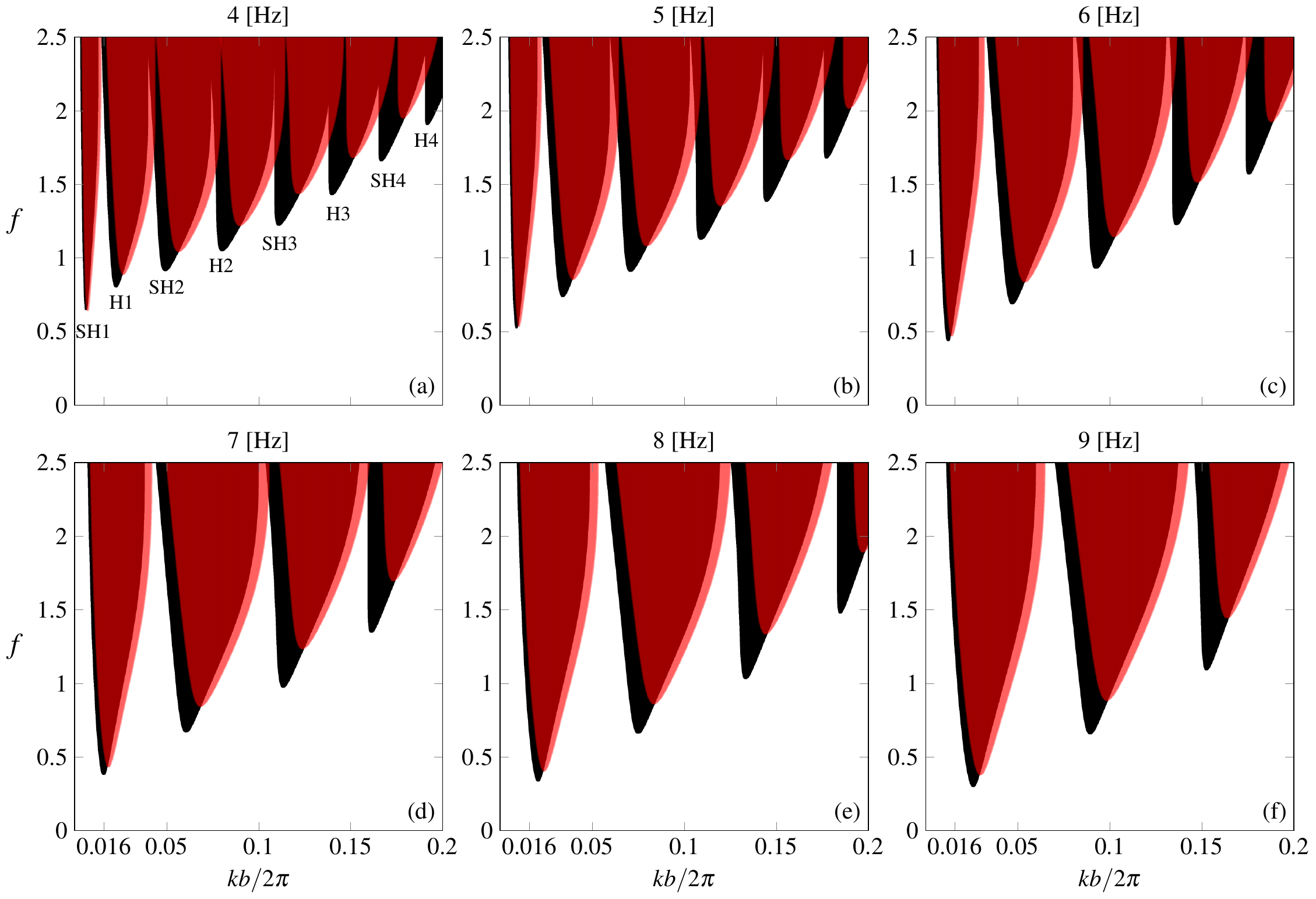}
\caption{Faraday tongues computed via Floquet analysis at different fixed driving frequencies (reported on the top of each panel). Black regions correspond to the unstable Faraday tongues computed using $\sigma_{BL}=12\nu/b^2$ as in the standard Darcy approximation, whereas red regions are the unstable tongues computed with the present modified $\sigma_{BL}=\chi_n\nu/b^2$. For this example, we consider ethanol 99.7\% (see table~\ref{tab:FluidProp}) in a Hele-Shaw cell of gap size $b=2\,\text{mm}$ filled to a depth $h=60\,\text{mm}$. $f$ denotes the non-dimensional forcing acceleration, $f=a\Omega^2/g$, with dimensional forcing amplitude $a$ and angular frequency $\Omega$. For plotting, we define a small scale-separation parameter $\epsilon= kb/2\pi$ and we arbitrarily set its maximum acceptable value to 0.2. Contact line dissipation is not included, i.e. $M=\sigma_{CL}=0$. SH stands for sub-harmonic, whereas $H$ stands for harmonic.}
\label{fig:C8_Faraday_tongues_comparison} 
\end{figure}

Given its periodic nature, the stability of the base flow, represented by a time-periodic modulation of the hydrostatic pressure, can be investigated via Floquet analysis. We therefore introduce the following Floquet ansatz \citep{kumar1994parametric}
\begin{subequations}
\begin{equation}
    \label{eq:C8_Floquet_ans_uvw}
    \mathbf{u}\left(x,y,z,t\right)=e^{\mu_F t}\sum_{n=-\infty}^{+\infty}\tilde{\mathbf{u}}_n\left(x,y,z\right) e^{\text{i}\left(n+\alpha/\Omega\right) t}=e^{\mu_F t}\sum_{n=-\infty}^{+\infty}\tilde{\mathbf{u}}_n\left(x,y,z\right) e^{\text{i}\xi_n t},
\end{equation}
\begin{equation}
    \label{eq:C8_Floquet_ans_p}
    p\left(x,z,t\right)=e^{\mu_F t}\sum_{n=-\infty}^{+\infty}\tilde{p}_n\left(x,z\right) e^{\text{i}\left(n+\alpha/\Omega\right) t}=e^{\mu_F t}\sum_{n=-\infty}^{+\infty}\tilde{p}_n\left(x,z\right) e^{\text{i}\xi_n t},
\end{equation}
\end{subequations}
\noindent where $\mu_F$ is the real part of the non-dimensional Floquet exponent and represents the growth rate of the perturbation. We have rewritten $\left(n+\alpha/\Omega\right)=\xi$ to better explicit the parametric nature of the oscillation frequency of the wave response. In the following, we will focus on the condition for marginal stability (boundaries of the Faraday's tongues), which require the growth rate $\mu_F=0$. In addition, values of $\alpha=0$ and $\Omega/2$ correspond, respectively, to harmonic and sub-harmonic parametric resonances \citep{kumar1994parametric}. This implies that $\xi$ is a parameter whose value is either $n$, for harmonics, or $n+1/2$, for sub-harmonics, with $n$ an integer $n=0,1,2,\hdots$. We will therefore use only a discrete version of $\xi$, namely $\xi_n$, with the index $n$ specific to each Fourier component in~\eqref{eq:C8_Floquet_ans_uvw}-\eqref{eq:C8_Floquet_ans_p}.\\
\indent  By injecting the ansatzs~\eqref{eq:C8_Floquet_ans_uvw}-\eqref{eq:C8_Floquet_ans_p} in~\eqref{eq:C8_NS_HS_nodim_uv}, we find that each component of the Fourier series must satisfy
\begin{equation}
        \label{eq:C8_NS_HS_nodim_uvn}
        \forall n: \ \ \ \ \ \text{i}\xi_n \tilde{u}_n=-\frac{\partial \tilde{p}_n}{\partial x}+\frac{\delta_{St}^2}{2} \frac{\partial^2 \tilde{u}_n}{\partial y^2},\ \ \ \ \ \text{i}\xi_n \tilde{w}_n=-\frac{\partial \tilde{p}_n}{\partial z}+\frac{\delta_{St}^2}{2} \frac{\partial^2 \tilde{w}_n}{\partial y^2},
\end{equation}
\noindent which, along with the no-slip condition at $y=\pm 1/2$, correspond to a two-dimensional pulsatile Poiseuille flow with solution
\begin{equation}
        \label{eq:C8_NS_HS_nodim_uvn_sol}
        \tilde{u}_n=\frac{\text{i}}{\xi_n}\frac{\partial \tilde{p}_n}{\partial x}\, F_n\left(y\right),\ \ \ \ \ \tilde{w}_n=\frac{\text{i}}{\xi_n}\frac{\partial \tilde{p}_n}{\partial z}\, F_n\left(y\right),\ \ \ \ \ F_n\left(y\right)=\left(1-\frac{\cosh{\left(1+\text{i}\right)y/\delta_n}}{\cosh{\left(1+\text{i}\right)/2\delta_n}}\right),
\end{equation}
\noindent and where $\delta_n=\delta_{St}/\sqrt{\xi_n}$, is a rescaled Stokes boundary layer thickness specific to the $n$th Fourier component. The function $F_n\left(y\right)$ is displayed in figure~\ref{fig:C8_profile_fz_chin}(b), which depicts how a decrease in the value of $\delta_n$ starting from large values corresponds to a progressive transition from a fully developed flow profile to a plug flow connected to thin boundary layers.\\
\indent  The gap-averaged velocity along the $y$-direction satisfies a Darcy-like equation,
\begin{equation}
\label{eq:C8_Darcy_like_eq}
<\tilde{\mathbf{u}}_n>=\int_{-1/2}^{1/2}\tilde{\mathbf{u}}_n\,\text{d}y=\frac{\text{i}\beta_n}{\xi_n}\nabla \tilde{p}_n,\ \ \ \ \ \ \ \ \beta_n=1-\frac{2\delta_n}{1+\text{i}}\tanh{\frac{1+\text{i}}{2\delta_n}}.
\end{equation}
\noindent \noindent In order to obtain a governing equation for the pressure $\tilde{p}_n$, we average the continuity equation and we impose the impermeability condition for the span-wise velocity, $v=0$ at $y=\pm 1/2$,
\begin{equation}
    \label{eq:C8_ave_cont_n}
    \frac{\partial <\tilde{u}_n>}{\partial x}+\underbrace{\int_{-1/2}^{1/2}\frac{\partial \tilde{v}_n}{\partial y}\,\text{d} y}_{v\left(1/2\right)-v\left(-1/2\right)=0}+\frac{\partial <\tilde{w}_n>}{\partial z}=\nabla\cdot<\tilde{\mathbf{u}}_n>=0,
\end{equation}
\noindent Since $<\tilde{\mathbf{u}}_n>=\text{i}\left(\beta_n/\xi_n\right) \nabla \tilde{p}_n$, the pressure field $\tilde{p}_n$ must obey the Laplace equation
\begin{equation}
    \label{eq:C8_Laplace_pn}
    \nabla^2 \tilde{p}_n=\frac{\partial^2 \tilde{p}_n}{\partial x^2}+\frac{\partial^2 \tilde{p}_n}{\partial z^2}=0.
\end{equation}
\noindent It is now useful to expand each Fourier component $\tilde{p}_n\left(x,z\right)$ in the infinite $x$-direction as $\sin{x}$ such that the $y$-average implies,
\begin{subequations}
\begin{equation}
    \label{eq:C8_ansatz_sinkx_pn}
    \tilde{p}_n=\hat{p}_n\left(z\right) \sin{x},
\end{equation}
    \begin{equation}
    \label{eq:C8_ansatz_sinkx_un_vn}
    <\tilde{u}_n>=\hat{u}_n=\frac{\text{i}\beta_n}{\xi_n} \hat{p}_n\cos{x},\ \ \ \ \ <\tilde{w}_n>=\hat{w}_n=\frac{\text{i}\beta_n}{\xi_n} \frac{\partial \hat{p}_n}{\partial z}\sin{x}.
\end{equation}
\end{subequations}
\noindent Replacing~\eqref{eq:C8_ansatz_sinkx_pn} in~\eqref{eq:C8_Laplace_pn} leads to 
\begin{equation}
    \label{eq:C8_Laplace_p_sinkx}
    \left(\frac{\partial^2 }{\partial z^2}-1\right)\hat{p}_n=0,
\end{equation}
\noindent which admits the solution form
\begin{equation}
    \label{eq:C8_Laplace_p_sinkx_sol}
    \hat{p}_n=c_{1}\cosh{z}+c_{2}\sinh{z}.
\end{equation}
\noindent The presence of a solid bottom imposes that $\hat{w}_n=0$ and, therefore, that $\partial \hat{p}_n/\partial z=0$, at a non-dimensional fluid depth $z=-hk$, hence giving 
\begin{equation}
    \label{eq:C8_press_sol}
    \hat{p}_n=c_{1}\left[\cosh{z}+\tanh{kh}\sinh{z}\right].
\end{equation}
\noindent Let us now invoke the linearized kinematic boundary condition
\begin{equation}
    \label{eq:C8_kin0}
    \frac{\partial \eta}{\partial t} =w.
\end{equation}
\noindent Note that free surface elevation, $\eta'\left(x',y',t'\right)$, has been rescaled by the forcing amplitude $a$, i.e. $\eta'/a=\eta$, and represents the projection of the bottom of the transverse concave meniscus on the $xz$-plane of figure~\ref{fig:C8_Sketch1}(a).
\noindent Moreover, by recalling the Floquet ansatzs~\eqref{eq:C8_Floquet_ans_uvw}-\eqref{eq:C8_Floquet_ans_p} (with $\mu_F=0$), here specified for the interface, we get an equation for each Fourier component $n$,
\begin{equation}
    \label{eq:C8_Floquet_ans_zeta}
    \eta=\sum_{n=-\infty}^{+\infty}\tilde{\eta}_n e^{\text{i}\xi_n t} \ \ \ \ \ \ \longrightarrow \ \ \ \ \ \ \forall n : \ \ \ \ \text{i}\xi_n\tilde{\eta}_n=\tilde{w}_n.
\end{equation}
\noindent Expanding $\tilde{\eta}_n$ in the $x$-direction as $\sin{x}$ and averaging in $y$, i.e. $<\tilde{\eta}_n>=\hat{\eta}_n\sin{x}$, leads to
\begin{equation}
    \label{eq:C8_averaged_kin}
    \text{i}\xi_n\hat{\eta}_n=\hat{w}_n=\frac{\text{i}\beta_n}{\xi_n}\frac{\partial \hat{p}_n\left(z=0\right)}{\partial z}=\frac{\text{i}\beta_n}{\xi_n} c_{1}\tanh{kh}\ \ \ \longrightarrow\ \ \ c_{1}=\frac{\xi_n^2}{\beta_n}\frac{\hat{\eta}_n}{\tanh{kh}}.
\end{equation}
\noindent Lastly, we consider the linearized dynamic condition (or linearized normal stress), evaluated at $z'=\eta'$ and where the term associated with the curvature of the free surface appears,
\begin{equation}
\label{eq:C8_dyn_bc}
-p'+\rho G\left(t'\right)\eta'+2\mu\frac{\partial w'}{\partial z'}-\gamma \left.\frac{\partial \kappa'}{\partial \eta'}\right|\eta'=0.
\end{equation}
\noindent In~\eqref{eq:C8_dyn_bc}, $\partial \kappa'/\partial \eta'$ represents the first-order variation of the curvature associated with the small perturbation $\eta'$. Capillary force in the $x$-direction is only important at large enough wavenumbers, although the associated term can be retained in the analysis in order to retrieve the dispersion relation for capillary-gravity waves \citep{li2019stability}. On the other hand, the small gap of Hele-Shaw cells is such that surface tension effects in the narrow $y$-direction are strongly exacerbated. In general, the curvature can be divided into two parts \citep{saffman1958penetration,chuoke1959instability}:
\begin{equation}
    \label{eq:C8_curv_terms}
     \kappa'\left(\eta'\right)=\frac{\partial}{\partial x'} \left(\frac{\partial_{x'} \eta'}{\sqrt{1+\left(\partial_{x'}\eta'\right)^2}}\right)+\frac{2}{b}\cos{\theta},
\end{equation}
\noindent where the first term indicates the principal radius of curvature and the second term represents the out-of-plane curvature of the meniscus (see figure~\ref{fig:C8_Sketch1}(a)). A common treatment of Hele-Shaw cells assumes the out-of-plane interface shape to be semicircular \citep{saffman1958penetration,mclean1981effect,park1984two,afkhami2013volume}. Nevertheless, laboratory observations have unveiled that liquid oscillations in Hele-Shaw cells experience an up-and-down driving force with $\theta$ constantly changing \citep{jiang2004contact}, hence giving rise to a dynamic contact angle. Here, as in \cite{li2019stability}, we use the following model \citep{hamraoui2000can} to evaluate the cosine of the dynamic contact angle $\theta$ as
\begin{equation}
    \label{eq:C8_contact_line_mod}
    \cos{\theta}=1-\frac{M}{\mu}Ca=1-\frac{M w'}{\gamma}
\end{equation}
\noindent where $Ca=\mu w'/\gamma$ is the Capillary number defined using the vertical contact line velocity $w'=\partial\eta'/\partial t'$. The friction coefficient $M$, sometimes referred to as mobility parameter $M$ \citep{xia2018moving}, can be interpreted in the framework of molecular kinetics theory \citep{voinov1976hydrodynamics,Hocking87,blake1993dynamic,blake2006physics,johansson2018molecular}, but here, in the same spirit of \cite{li2019stability}, we simply view this coefficient as a constant phenomenological fitting parameter that defines the energy dissipation rate per unit length of the contact line.\\
\indent  By combining equations~\eqref{eq:C8_curv_terms}-\eqref{eq:C8_contact_line_mod} and taking their first-order curvature variation applied to the small perturbation, one can express
\begin{equation}
\label{eq:C8_firstordkappa}
-\gamma\left.\frac{\partial\kappa'}{\partial \eta'}\right|\eta'=-\gamma\frac{\partial^2\eta'}{\partial x'^2}+\frac{2M}{b}\frac{\partial \eta'}{\partial t'}.
\end{equation}
\noindent After turning to non-dimensional quantities using the scaling in~\eqref{eq:C8_scaling_nondim}, equations~\eqref{eq:C8_dyn_bc} reads
\begin{equation}
\label{eq:C8_dyn_nondim}
-\Omega^2p+g\eta-\frac{\gamma}{\rho}k^2\frac{\partial ^2\eta}{\partial x^2}+\frac{2M}{\rho b}\Omega \frac{\partial \eta}{\partial t}=a\Omega^2\cos{\Omega t'}\eta,
\end{equation}
\noindent where the viscous stress term has been eliminated, as it is negligible compared to the others.\\
\indent  With introduction of the Floquet ansatz~\eqref{eq:C8_Floquet_ans_p}-\eqref{eq:C8_Floquet_ans_zeta} and by recalling the $x$-expansion of the interface and pressure as $\sin{x}$, the averaged normal stress equations become
\begin{equation}
\label{eq:C8_aver_stress}
\forall n:\ \ \ \ -\Omega^2\hat{p}_n+\text{i}\left(\xi_n\Omega\right)\frac{2M}{\rho b}\hat{\eta}_n+\left(1+\frac{\gamma}{\rho g}k^2\right)g\hat{\eta}_n=\frac{a\Omega^2}{2g}g\left(\hat{\eta}_{n-1}+\hat{\eta}_{n+1}\right).
\end{equation}
\noindent where the decomposition $\cos{\Omega t'}=\left(e^{\text{i} \Omega t'}+e^{-\text{i}\Omega t'}\right)/2=\left(e^{\text{i} t}+e^{-\text{i} t}\right)/2$ has also been used. Equations~\eqref{eq:C8_press_sol} and~\eqref{eq:C8_averaged_kin} are finally used to express the dynamic equation as a function of the non-dimensional averaged interface only,
\begin{equation}
    \label{eq:C8_shear_y}
    -\frac{\left(\xi_n\Omega\right)^2}{\beta_n} \hat{\eta}_n +\text{i}\left(\xi_n\Omega\right)\frac{2M}{\rho b}k\tanh{kh}\hat{\eta}_n + \left(1+\Gamma\right)gk\tanh{kh}\,\hat{\eta}_n=\frac{gk\tanh{kh}}{2}\, f\left(\hat{\eta}_{n-1}+\hat{\eta}_{n+1}\right),
\end{equation}
\noindent with the auxiliary variables $f=a\Omega^2/g$ and $\Gamma=\gamma k^2/\rho g$, such that $\left(1+\Gamma\right)gk\tanh{kh}=\omega_0^2$, the well-known dispersion relation for capillary-gravity waves \citep{Lamb32}.\\
\indent  As in the present form the interpretation of coefficient $\beta_n$ does not appear straightforward, it is useful to define the damping coefficients
\begin{subequations}
\begin{equation}
    \label{eq:C8_damping_coeffs}
    \sigma_n=\sigma_{BL}+\sigma_{CL},\ \ \ \ \ \sigma_{BL}=\chi_n\frac{\nu}{b^2},\ \ \ \ \ \sigma_{CL}=\frac{2M}{\rho b}k\tanh{kh},
\end{equation}
\noindent where $\chi_n$ is used to help rewriting $\frac{1}{\beta_n}=1-\text{i}\frac{\delta_n^2}{2}\chi_n$,
\begin{equation}
    \label{eq:C8_damping_BL_chi}
    \chi_n=\text{i}\frac{2}{\delta_n^2}\left(\frac{1-\beta_n}{\beta_n}\right)=12\left[\frac{\text{i}}{6\delta_n^2}\left( \frac{\frac{2\delta_n}{1+\text{i}}\tanh{\frac{1+\text{i}}{2\delta_n}}}{1-\frac{2\delta_n}{1+\text{i}}\tanh{\frac{1+\text{i}}{2\delta_n}}} \right)\right].
\end{equation}
\end{subequations}
\noindent These auxiliary definitions allows one to express~\eqref{eq:C8_shear_y} as
\begin{equation}
    \label{eq:C8_semi_final}
    -\left(\xi_n\Omega\right)^2 \hat{\eta}_n +\text{i}\left(\xi_n\Omega\right)\sigma_n \hat{\eta}_n + \omega_0^2\hat{\eta}_n= \frac{\omega_0^2}{2\left(1+\Gamma\right)}\, f\left[\hat{\eta}_{n+1}+\hat{\eta}_{n-1}\right)].
\end{equation}
\noindent or, equivalently,
\begin{equation}
    \label{eq:C8_Mathieu}
    \frac{2\left(1+\Gamma\right)}{\omega_0^2}\left[-\left(n\Omega+\alpha\right)^2+\text{i}\left(n\Omega+\alpha\right)\sigma_n+\omega_0^2\right]\hat{\eta}_n=f \left[\hat{\eta}_{n+1}+\hat{\eta}_{n-1}\right].
\end{equation}
\noindent Subscripts $BL$ and $CL$ in~\eqref{eq:C8_damping_coeffs} denote, respectively, the boundary layers and contact line contributions to the total damping coefficient $\sigma_n$. 
\\
\indent  At the end of this long mathematical derivation, the main result is the modified damping coefficient $\sigma_n$. Since the boundary layer contribution, $\sigma_{BL}$ depends on the $n$th Fourier component, the overall damping, $\sigma_n$, is mode dependent and its value is different for each specific $n$th parametric resonant tongue considered. This is in stark contrast with the standard Darcy approximation, where $\sigma_{BL}$ is the same for each resonance and amounts to $12\nu/b^2$. In our model, the case of $\alpha=0$ with $n=0$ constitutes a peculiar case, as $\xi_n=\xi_0=0$ and $\delta_0\rightarrow+\infty$. In such a situation, $F_0\left(y\right)$ tends to the steady Poiseuille profile, so that we take $\chi_0=12$.\\
\indent  Similarly to \cite{kumar1994parametric}, equation~\eqref{eq:C8_Mathieu} is rewritten as
\begin{equation}
    \label{eq:C8_Mathieu1}
    A_n\hat{\eta}_n=f\left[\hat{\eta}_{n+1}+\hat{\eta}_{n-1}\right],
\end{equation}
\noindent with
\begin{equation}
    \label{eq:C8_AnC}
    A_n=\frac{2\left(1+\Gamma\right)}{\omega_0^2}\left(-\left(n\Omega+\alpha\right)^2+\text{i}\left(n\Omega+\alpha\right)\sigma_n+\omega_0^2\right)=A_n^r+\text{i}A_n^i\in\mathbb{C}
\end{equation}
\noindent The non-dimensional amplitude of the external forcing, $f=a\Omega^2/g$ appears linearly, therefore~\eqref{eq:C8_Mathieu1} can be considered to be a generalized eigenvalue problem
\begin{equation}
    \label{eq:C8_Mathieu1_eig}
    \mathbf{A}\hat{\mathbf{\eta}}=f\mathbf{B}\hat{\mathbf{\eta}},
\end{equation}
\noindent with eigenvalues $f$ and eigenvectors whose components are the real and imaginary parts of $\hat{\eta}_n$. See \cite{kumar1994parametric} for the structure of matrices $\mathbf{A}$ and $\mathbf{B}$.\\
\indent  For one frequency forcing we use a truncation number $N=10$, which produces $2\left(N+1\right)\times 2\left(N+1\right) = 22\times 22$ matrices. Eigen-problem~\eqref{eq:C8_Mathieu1_eig} is then solved in Matlab using the built-in function \textit{eigs}, by asking for the eigenvalue (or few eigenvalues) with the smallest real part.\\
\indent Figure~\ref{fig:C8_Faraday_tongues_comparison} shows the results of this procedure for one of the configurations considered by \cite{li2019stability} and neglecting the dissipation associated with the contact line motion, i.e. $M=0$. In each panel, associated with a fixed forcing frequency, the black regions correspond to the unstable Faraday tongues computed using $\sigma_{BL}=12\nu/b^2$ as given by Darcy's approximation, whereas the red regions are the unstable tongues computed with the modified $\sigma_{BL}=\chi_n\nu/b^2$. At a forcing frequency $4\,\text{Hz}$ the first sub-harmonic tongues computed using the two models essentially overlap. Yet, successive resonances display an increasing departure from Darcy's model due to the newly introduced complex coefficient $\sigma_n$. Particularly, the real part of $\chi_n$ is responsible for the higher onset acceleration, while the imaginary part is expected to act as a detuning term, which shifts the resonant wavenumbers $k$.

\subsection{Asymptotic approximations}\label{subsec:C8_Sec1sub5}

The main result of this analysis consists in the derivation of the modified damping coefficient $\sigma_n=\sigma_{n,r}+\text{i}\sigma_{n,i}$ associated with each parametric resonance. Aiming at better elucidating how this modified complex damping influences the stability properties of the system, we would like to derive in this section an asymptotic approximation, valid in the limit of small forcing amplitudes, damping and detuning, of the first sub-harmonic (SH1) and harmonic (H1) Faraday tongues.\\
\indent  Unfortunately, the dependence of $\sigma_n$ on the parametric resonance considered and, more specifically, on the $n$th Fourier component, does not allow one to convert~\eqref{eq:C8_semi_final}, expressed in a discrete frequency domain, back into the continuous temporal domain. By keeping this in mind, we can still imagine fixing the value of $\sigma_n$ to that corresponding to the parametric resonance of interest, e.g. $\sigma_0$ (with $n=0$ and $\xi_0\Omega=\Omega/2$) for SH1 or $\sigma_1$ (with $n=1$ and $\xi_1\Omega=\Omega$) for H1. By considering then that for the SH1 and H1 tongues, the system responds in time as $\text{exp}\left(\text{i}\Omega t/2\right)$ and $\text{exp}\left(\text{i}\Omega t \right)$, respectively, we can recast, for these two specific cases, equations~\eqref{eq:C8_semi_final} into a damped Mathieu equation \citep{Benjamin54,kumar1994parametric,muller1997analytic}
\begin{equation}
    \label{eq:C8_Damp_Mathieu_fin}
    \frac{\partial^2 \hat{\eta}}{\partial t'^2}+\hat{\sigma}_{n}\frac{\partial \hat{\eta}}{\partial t'}+\omega_0^2\left(1-\frac{f}{1+\Gamma}\cos{\Omega t'}\right)\hat{\eta}=0.
\end{equation}
\noindent with either $\hat{\sigma}_{n}=\sigma_0$ (SH1) or $\hat{\sigma}_n=\sigma_1$ (H1) and where one can recognize that $-\left(\xi_n\Omega\right)^2\hat{\eta}\leftrightarrow\partial^2\hat{\eta}/\partial t'^2$ and $\text{i}\left(\xi_n\Omega\right)\hat{\eta}\leftrightarrow\partial \hat{\eta}/\partial t'$. Asymptotic approximations can be then computed by expanding asymptotically the interface as $\hat{\eta}=\hat{\eta}_0+\epsilon\hat{\eta}_1+\epsilon^2\hat{\eta}_2+\hdots$, with $\epsilon$ a small parameter $\ll 1$.

\subsubsection{First sub-harmonic tongue}\label{subsubsec:C8_Sec1sub5subsub1}

\begin{figure}
\centering
\includegraphics[width=0.85\textwidth]{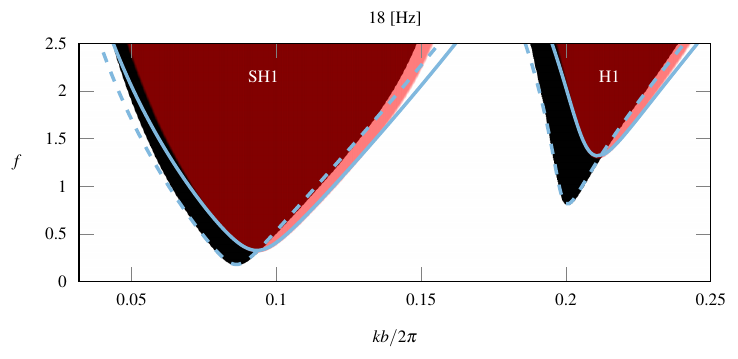}
\caption{First sub-harmonic and harmonic Faraday tongues at a driving frquency $1/T=18\,\text{Hz}$ for the same configuration of figure~\ref{fig:C8_Faraday_tongues_comparison}. Black and red regions show unstable tongues computed via Floquet analysis by using, respectively, $\sigma_{BL}=12\nu/b^2$ and the modified $\sigma_{BL}=\chi_1\nu/b^2$ from the present model. Dashed and solid light-blue lines correspond to the asymptotic approximations according to \eqref{eq:C8_asym_SH} and~\eqref{eq:C8_asym_H_rel}.}
\label{fig:C8_comp_WNL} 
\end{figure}

As anticipated above, when looking at the first or fundamental sub-harmonic tongue (SH1), one should take $\hat{\sigma}_n\rightarrow\sigma_{0}$ (with $\xi_{0}\Omega=\Omega/2$), which is assumed small of order $\epsilon$. The forcing amplitude $f$ is assumed of order $\epsilon$ as well. Furthermore, a small detuning $\sim\epsilon$, such that $\Omega=2\omega_0+\epsilon\lambda$, is also considered, and, in the spirit of the multiple timescale analysis, a slow time scale $T=\epsilon t'$ \citep{Nayfeh2008} is introduced. At leading order, the solution reads $\hat{\eta}_0=A\left(T\right)e^{\text{i}\omega_0 t'}+c.c.$, with $c.c.$ denoting the complex conjugate part. At the second order in $\epsilon$, the imposition of a solvability condition necessary to avoid secular terms prescribes the amplitude $B\left(T\right)=A\left(T\right)e^{-\text{i}\lambda T/2}$ to obey the following amplitude equation
\begin{equation}
    \label{eq:C8_amp_eq_sub_harm_rect}
    \frac{d B}{d T}=-\frac{\sigma_{0}}{2}B-\text{i}\frac{\lambda}{2}B-\text{i}\frac{\omega_0}{4\left(1+\Gamma\right)}f \overline{B}.
\end{equation}
\noindent Turning to polar coordinates, i.e. $B=|B|e^{\text{i}\Phi}$, keeping in mind that $\sigma_{0}=\sigma_{0,r}+\text{i}\sigma_{0,i}$ and looking for stationary solutions with $|B|\ne 0$ (we skip the straightforward mathematical steps), one ends up with the following approximation for the marginal stability boundaries associated with the first sub-harmonic Faraday tongue
\begin{equation}
    \label{eq:C8_asym_SH}
    \left(\frac{\Omega+\sigma_{0,i}}{2\omega_0}-1\right)=\pm\frac{1}{4\left(1+\Gamma\right)}\sqrt{f^2-\frac{4\sigma_{0,r}^2\left(1+\Gamma\right)^2}{\omega_0^2}},
\end{equation}
\noindent whose onset acceleration value, $\min{f_{1_{SH}}}$, amounts to
\begin{equation}
    \label{eq:C8_asym_SH_min_th}
    \min{f_{SH1}} = 2\sigma_{0,r}\sqrt{\frac{1+\Gamma}{gk\tanh{kh}}}\approx 2\sigma_{0,r}\sqrt{\frac{1}{g}\left(\frac{1}{k}+\frac{\gamma}{\rho g}k \right)},
\end{equation}
\noindent Note that the final approximation on the right-hand-side of~\eqref{eq:C8_asym_SH_min_th} only holds if $kh\gg1$, so that $\tanh{kh}\approx 1$ (deep water regime). Given that $\chi_{0,r}>12$ and $\chi_{0,i}>0$ always, the asymptotic approximation~\eqref{eq:C8_asym_SH_min_th}, in its range of validity, suggests that Darcy's model underestimates the sub-harmonic stability threshold. Moreover, from~\eqref{eq:C8_asym_SH}, the critical wavenumber $k$, associated with $\min{f_{SH1}}$, would correspond to that prescribed by the Darcy approximation but at an effective forcing frequency $\Omega+\sigma_{0,i}=2 \omega_0$ instead of at $\Omega=2\omega_0$. This explains why the modified tongues appear shifted towards higher wavenumbers. These observations are well visible in figure~\ref{fig:C8_comp_WNL}. 

\subsubsection{First harmonic tongue}\label{subsubsec:C8_Sec1sub5subsub2}

By analogy with \S\ref{subsubsec:C8_Sec1sub5subsub1}, an analytical approximation of the first harmonic tongue (H1) can be provided. In the same spirit of \cite{rajchenbach2015faraday}, we adapt the asymptotic scaling such that $f$ is still of order $\epsilon$, but $T=\epsilon^2$, $\hat{\sigma}_n=\sigma_1\sim\epsilon^2$ (with $\xi_{1}\Omega=\Omega$) and $\Omega=\omega_0+\epsilon^2\lambda$. Pursuing the expansion up to $\epsilon^2$-order, with $\hat{\eta}_0=A\left(T\right)e^{\text{i}\omega_0 t'}+c.c.$ and $B\left(T\right)=A\left(T\right)e^{-\text{i}\lambda T}$, will provide the amplitude equation  
\begin{equation}
    \label{eq:C8_amp_eq_harm_rect}
    \frac{d B}{d T}=-\frac{\sigma_{1}}{2}B-\text{i}\lambda B-\text{i}\frac{\omega_0}{8\left(1+\Gamma\right)^2}f^2 \overline{B}+\text{i}\frac{\omega_0}{12\left(1+\Gamma\right)^2}f^2 B.
\end{equation}
\noindent The approximation for the marginal stability boundaries derived from~\eqref{eq:C8_amp_eq_harm_rect} takes the form
\begin{equation}
    \label{eq:C8_asym_H_rel}
    \left(\frac{\Omega+\sigma_{1,i}/2}{\omega_0}-1\right)=\frac{f^2}{12\left(1+\Gamma\right)^2}\pm\frac{1}{8\left(1+\Gamma\right)^2}\sqrt{f^4-\left(\frac{4\sigma_{1,r}\left(1+\Gamma\right)^2}{\omega_0}\right)^2}
\end{equation}
\noindent with a minimum onset acceleration, $\min{f_{1_{H}}}$
\begin{equation}
    \label{eq:C8_asym_H_min_th}
    \min{f_H} = 2\sqrt{\sigma_{1,r}}\left(\frac{\left(1+\Gamma\right)^3}{gk\tanh{kh}}\right)^{1/4}\approx 2\sqrt{\sigma_{1,r}}\frac{1}{g^{1/4}}\left(\frac{1}{k^{1/3}}+\frac{\gamma}{\rho g}k^{5/3}\right)^{3/4},
\end{equation}
\noindent and where, as before, the final approximation on the right-hand side is only valid in the deep water regime. Similarly to the sub-harmonic case, the critical wavenumber $k$ corresponds to that prescribed by the Darcy approximation but at an effective forcing frequency $\Omega+\sigma_{1,i}/2=\omega_0$ instead of at $\Omega=\omega_0$ and the onset acceleration is larger than that predicted from the Darcy approximation (as $\chi_{1,r}>12$).

\subsection{Comparison with experiments by \cite{li2019stability}}\label{subsec:C8_Sec1sub4}

\begin{table}
\centering
\begin{tabular}{ccccccccccccccccc}
Liquid & & & & $\mu$ $\left[\text{mPa s}\right]$ & & & & $\rho$ $\left[\text{kg/m$^3$}\right]$ & & & & $\gamma$ $\left[\text{N/m}\right]$ & & & & $M$ $\left[\text{Pa s}\right]$\\ \hline
ethanol 99.7\% & & & & 1.096 & & & & 785 & & & & 0.0218 & & & & 0.04\\
ethanol 70.0\% & & & & 2.159 & & & & 835 & & & & 0.0234 & & & & 0.0485\\
ethanol 50.0\% & & & & 2.362 & & & & 926 & & & & 0.0296 & & & & 0.07\\
\end{tabular}
\caption{Characteristic fluid parameters for the three ethanol-water mixtures considered in this study. Data for the pure ethanol and ethanol-water mixture (50\%) are taken from \cite{li2019stability}. The value of the friction parameter $M$ for ethanol-70\% is fitted from the experimental measurements reported in \S\ref{sec:C8_EXP}, but lies well within the range of values used by \cite{li2019stability} and agrees with the linear trend displayed in figure~5 of \cite{hamraoui2000can}.}
\label{tab:FluidProp}
\end{table}
\begin{figure}
\centering
\includegraphics[width=1\textwidth]{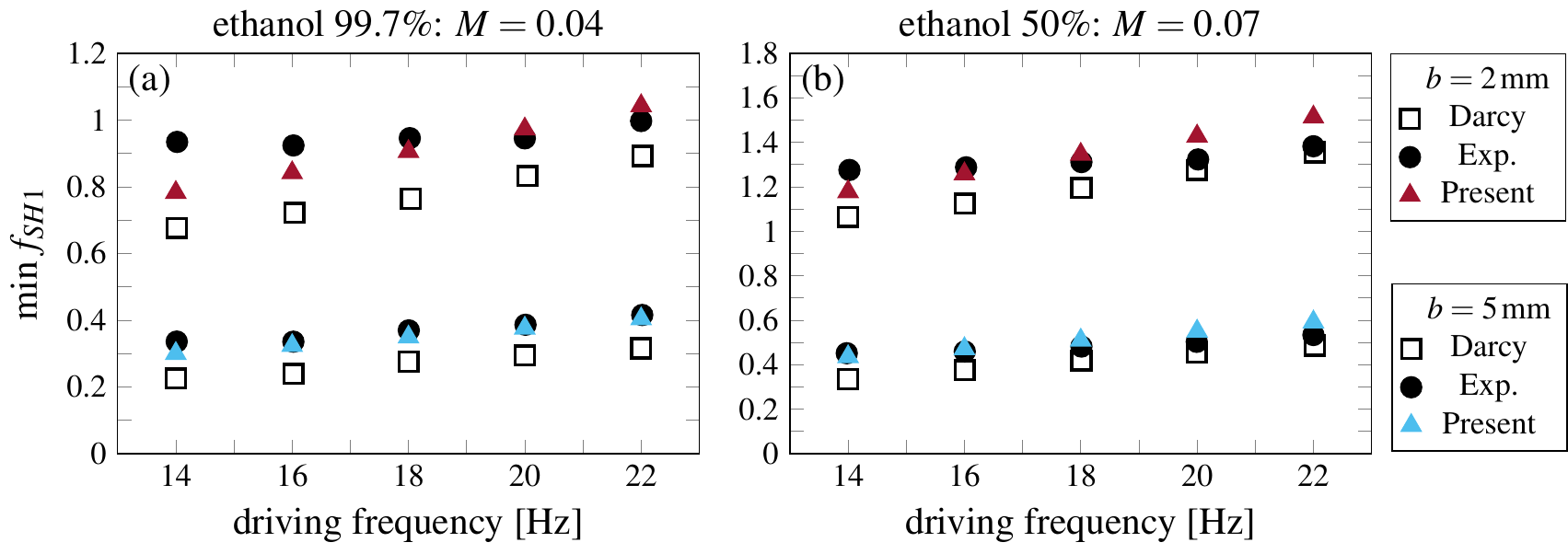}
\caption{Sub-harmonic instability onset, $\min{f}$, versus forcing frequency. Comparison between theoretical data (empty squares: standard Darcy model, $\sigma_{BL}=12\nu/b^2$; colored triangles: present model, $\sigma_{BL}=\chi_n\nu/b^2$) and experimental measurements by \cite{li2019stability}. The values of the mobility parameter $M$ here employed are reported in the figure.}
\label{fig:C8_comp_exp} 
\end{figure}

Results presented so far were produced by assuming the absence of contact line dissipation, i.e. coefficient $M$ was set to $M=0$, so that $\sigma_{CL}=0$. In this section, we reintroduce such a dissipative contribution and we compare our theoretical predictions with a set of experimental measurements reported by \cite{li2019stability}, using the values they have proposed for M. This comparison, shown in figure~\ref{fig:C8_comp_exp}, is outlined in terms of non-dimensional minimum onset acceleration, $\min{f}=\min{f_{SH1}}$, versus driving frequency. These authors performed experiments in two different Hele-Shaw cells of length $l=300\,\text{mm}$, fluid depth $h=60\,\text{mm}$ and gap-size $b=2\,\text{mm}$ or $b=5\,\text{mm}$. Two fluids, whose properties are reported in table~\ref{tab:FluidProp}, were used: ethanol 99.7\% and ethanol 50\%. The empty squares in figure~\ref{fig:C8_comp_exp} are computed via Floquet stability analysis~\eqref{eq:C8_Mathieu1_eig} using the Darcy approximation for $\sigma_{BL}=12\nu/b^2$ and correspond to the theoretical prediction by \cite{li2019stability}, while the colored triangles are computed using the corrected $\sigma_{BL}=\chi_n\nu/b^2$. Although the trend is approximately the same, the Darcy approximation underestimates the onset acceleration with respect to the present model, which overall compares better with the experimental measurements (black-filled circles). Some disagreement still exists, especially at smaller cell gaps, i.e. $b=2\,\text{mm}$, where surface tension effects are even larger. This is likely attributable to an imperfect phenomenological contact line model \citep{bongarzone2021relaxation,bongarzone2022sub}, whose definition falls beyond the scope of this work. Yet, this comparison shows how the modifications introduced by the present model contribute to closing the gap between theoretical Faraday onset estimates and these experiments.  

\section{The case of thin annuli}\label{sec:C8_Sec2}

 We now consider the case of a thin annular container, whose nominal radius is $R$ and the actual inner and outer radii are $R-b/2$ and $R+b/2$, respectively (see the sketch in figure~\ref{fig:C8_Sketch1}(b)). In the limit of $b/R\ll 1$, the wall curvature is negligible and the annular container can be considered a Hele-Shaw cell. The following change of variable for the radial coordinate, $r'=R+y'=R\left(1+y'/R\right)$ with $y'\in\left[-b/2,b/2\right]$, will be useful in the rest of the analysis. As in~\S\ref{sec:C8_Sec1}, we first linearize around the rest state. Successively, we introduce the following non-dimensional quantities,
\begin{equation}
    \label{eq:C8_nondim_annular}
    r=\frac{r'}{R},\ \ \ y=\frac{y'}{b},\ \ \ z=\frac{z'}{R},\ \ \ u=\frac{u_{\varphi}'}{a\Omega},\ \ \ v=\frac{u_r'}{a\Omega\left(b/R\right)},\ \ \ w=\frac{u_z'}{a\Omega^2},\ \ \ p=\frac{p'}{\rho Ra\Omega^2}.
\end{equation}
\noindent It follows that, at leading order, $r=R\left(1+yb/R\right)\sim R\longrightarrow1/r=1/\left(R\left(1+yb/R\right)\right)\sim 1/R$ but $\partial/\partial_r=\left(R/b\right)\partial/\partial_y\sim b/R\gg 1$. With this scaling and introducing the Floquet ansatzs~\eqref{eq:C8_Floquet_ans_uvw}-\eqref{eq:C8_Floquet_ans_p}, one obtains the following simplified governing equations,
\begin{subequations}
    \begin{equation}
        \label{eq:C8_ann_continuity}
        \frac{\partial \tilde{u}_n}{\partial \varphi}+\frac{\partial \tilde{v}_n}{\partial y}+\frac{\partial \tilde{w}_n}{\partial z}=0,
    \end{equation}
    \begin{equation}
        \label{eq:C8_ann_moment}
        \text{i}\tilde{u}_n=-\frac{1}{\xi_n}\frac{\partial \tilde{p}_n}{\partial \varphi}+\frac{\delta_{n}^2}{2}\frac{\partial ^2 \tilde{u}_n}{\partial y^2},\ \ \ \ \text{i} \tilde{w}_n=-\frac{1}{\xi_n}\frac{\partial \tilde{p}_n}{\partial z}+\frac{\delta_{n}^2}{2}\frac{\partial ^2 \tilde{w}_n}{\partial y^2}\ \ \ \ \text{or}\ \ \ \ \tilde{\mathbf{u}}_n=\frac{\text{i}}{\xi_n}\nabla \tilde{p}_n F_n\left(y\right),
    \end{equation}
\end{subequations}
\noindent which are fully equivalent to those for the case of conventional rectangular cells if the transformation $\varphi\rightarrow x$ is introduced. Averaging the continuity equation with the imposition of the no-penetration condition at $y=\mp 1/2$, $v\left(\mp 1/2\right)$, eventually leads to 
\begin{equation}
    \label{eq:C8_laplace_p_ann}
    \nabla^2 \tilde{p}_n=\frac{\partial^2 \tilde{p}_n}{\partial z^2}+\frac{\partial^2 \tilde{p}_n}{\partial \varphi^2},
\end{equation}
\noindent identically to~\eqref{eq:C8_Laplace_pn}. Expanding $\tilde{p}_n$ in the azimuthal direction as $\tilde{p}_n=\hat{p}_n\sin{m\varphi}$, with $m$ the azimuthal wavenumber, provides
\begin{equation}
    \label{eq:C8_laplace_p_ann_sin}
    \left(\frac{\partial^2}{\partial z^2}-m^2\right)\hat{p}_n=0\ \ \ \longrightarrow\ \ \ \hat{p}_n=c_{1}\cosh{mz}+c_{2}\sinh{mz},
\end{equation}
\noindent and the no-penetration condition at the solid bottom located at $z=-h/R$, $\hat{w}_n=\partial_z\hat{p}_n=0$, prescribes
\begin{equation}
    \label{eq:C8_laplace_p_ann_sol}
\hat{p}_n=c_{1}\left(\cosh{mz}+\tanh{mh/R}\sinh{mz}\right).
\end{equation}
\noindent Although so far the rectangular and the annular cases are indistinguishable from each other, here it is crucial to observe that the axisymmetric container geometry translates into a periodicity condition according to which 
\begin{equation}
    \label{eq:C8_periodicity_cond}
    \sin{\left(-m\pi\right)}=\sin{\left(m\pi\right)}\ \ \ \longrightarrow\ \ \ \sin{m\pi}=0,
 \end{equation}
\noindent and that always imposes the azimuthal wavenumber to be an integer. In other words, in contradistinction with the case of~\S\ref{sec:C8_Sec1}, where the absence of lateral wall ideally allows for any wavenumber $k$, here we have $m=0,1,2,3,\hdots\in\mathbb{N}$.\\
\indent  By repeating the calculations outlined in \S\ref{sec:C8_Sec1}, one ends up with the very same equation~\eqref{eq:C8_Mathieu} (and subsequent~\eqref{eq:C8_Mathieu1}-\eqref{eq:C8_Mathieu1_eig}), but where $\omega_0$ obeys to the \textit{quantized} dispersion relation
\begin{equation}
    \label{eq:C8_disp_rel_quant}
    \omega_0^2=\left(\frac{g}{R}m+\frac{\gamma}{\rho R^3}m^3\right)\tanh{m \frac{h}{R}}=\left(1+\Gamma\right)\frac{g}{R}m\tanh{m \frac{h}{R}}.
\end{equation}
\noindent with $\Gamma=\gamma m^2/\rho g R^2$. In this context, a representation of Faraday's tongues in the forcing frequency-amplitude plane appears most natural, as each parametric tongue will correspond to a fixed wavenumber $m$. Consequently, instead of fixing $\Omega$ and varying the wavenumber, here we solve~\eqref{eq:C8_Mathieu1_eig} by fixing $m$ and varying $\Omega$.

\subsection{Floquet analysis and asymptotic approximation}\label{subsec:C8_Sec2sub1}

\begin{figure}
\centering
\includegraphics[width=1\textwidth]{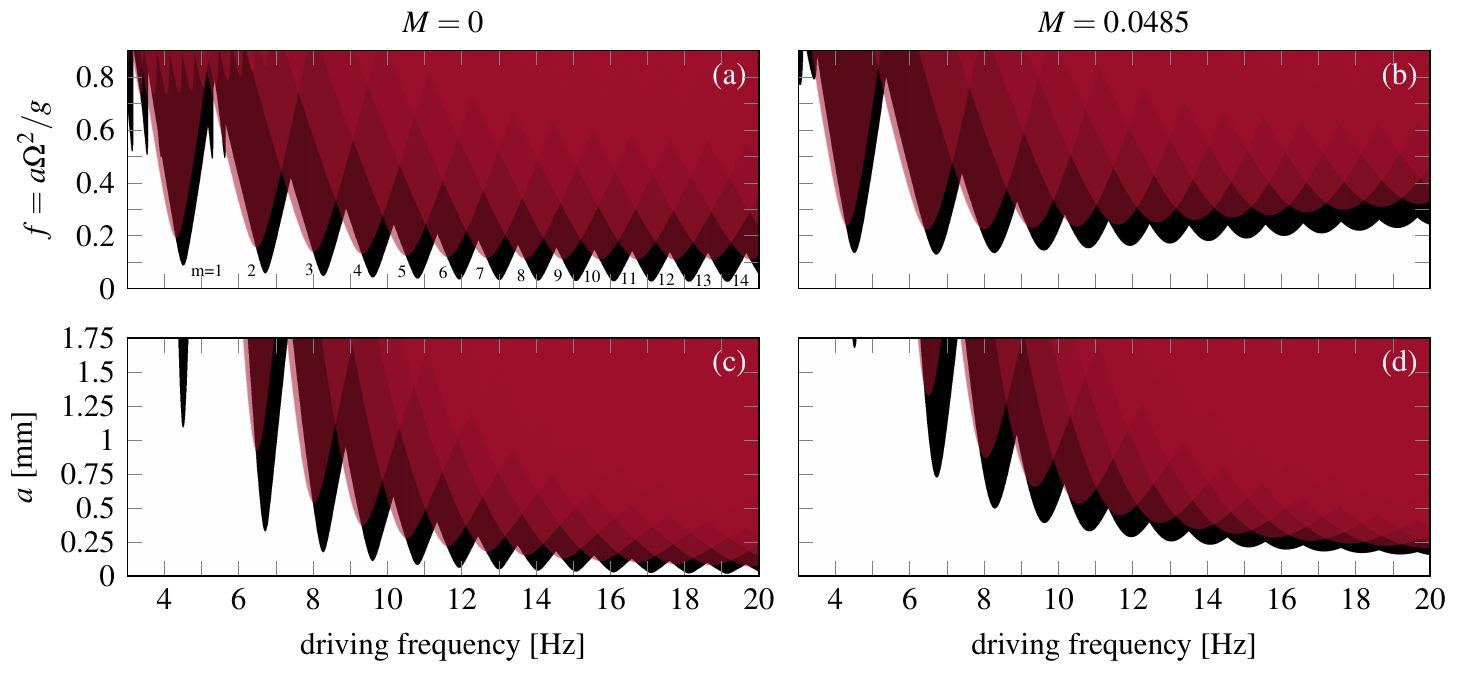}
\caption{Faraday tongues computed via Floquet analysis~\eqref{eq:C8_Mathieu1_eig} at different fixed azimuthal wavenumber $m$ and varying the driving frequency. (a)-(b) Faraday thresholds in terms of forcing acceleration $f=a\Omega^2/g$; (c)-(d) Threshold in terms of forcing amplitude $a$. Black regions correspond to the unstable Faraday tongues computed using $\sigma_{BL}=12\nu/b^2$, whereas red regions are the unstable tongues computed with the present modified $\sigma_{BL}=\chi_n\nu/b^2$. The fluid parameters used here correspond to those given in table~\ref{tab:FluidProp} for ethanol 70\%. The gap-size is set to $b=7\,\text{mm}$, the fluid depth to $h=65\,\text{mm}$ and the nominal radius to $R=44\,\text{mm}$. Contact line dissipation is included in (b) and (d) by accounting for a mobility coefficient $M=0.0485$. The regions with the lowest thresholds in each panel are sub-harmonic tongues associated with modes from $m=1$ to $14$.}
\label{fig:C8_Fig_WNL_Tongues_m} 
\end{figure}

The results from this procedure are reported in figure~\ref{fig:C8_Fig_WNL_Tongues_m}, where, as in figure~\ref{fig:C8_Faraday_tongues_comparison}, the black regions correspond to the unstable tongues obtained according to the standard gap-averaged Darcy model, while the red ones are computed using the corrected $\sigma_{BL}=\chi_n\nu/b^2$. The Faraday threshold is represented in terms of forcing acceleration (panels (a) and (b)) and forcing amplitude (panels (c) and (d)). In figure~\ref{fig:C8_Fig_WNL_Tongues_m}(a)-(c) no contact line model is included, whereas in (b)-(d) a mobility parameter $M=0.0485$ is accounted for. This specific value for $M$ will be used in the next section when comparing the theory with dedicated experiments. The regions with the lowest thresholds in each panel are sub-harmonic tongues associated with modes from $m=1$ to $14$. For the case with $M\ne0$, the instability onset acceleration associated with each wavenumber $m$ appears to follow a linear trend, as already reported in figure~\ref{fig:C8_comp_exp}.\\
\indent  In general, the present model gives a higher instability threshold, consistent with the results reported in the previous section. However, the tongues are here shifted to the left. This apparent opposite correction is a natural consequence of the different representations: varying wavenumber at a fixed forcing frequency (as in figure~\ref{fig:C8_Faraday_tongues_comparison}) versus varying forcing frequency at a fixed wavenumber (figure~\ref{fig:C8_Fig_WNL_Tongues_m}).\\
\indent  The asymptotic approximation for the sub-harmonic onset acceleration, adapted to this case from~\eqref{eq:C8_asym_SH} in \S\ref{subsubsec:C8_Sec1sub5subsub1},
\begin{equation}
    \label{eq:C8_asym_SH_m}
    f_{SH1} = 2 \sqrt{\left(1+\Gamma\right)\frac{\sigma_{0,r}^2}{\left(g/R\right)m\tanh{m h/R}}+4\left(1+\Gamma\right)^2\left(\frac{\Omega+\sigma_{0,i}}{2\omega_0}-1\right)^2},
\end{equation}
\noindent with
\begin{equation}
    \label{eq:C8_asym_SH_min_th_m}
    \min{f_{SH1}} = 2\sigma_{0,r}\frac{1+\Gamma}{\omega_0}=2\sigma_{0,r}\sqrt{\frac{1+\Gamma}{\left(g/R\right)m\tanh{m h/R}}}\approx 2\sigma_{0,r}\sqrt{\frac{R}{g}\left(\frac{1}{m}+\frac{\gamma}{\rho g R^2}m\right)},
\end{equation}
\noindent helps us indeed in rationalizing the influence of the modified complex damping coefficient.\\
\indent  The apparent opposite shift shown in figure~\ref{fig:C8_Fig_WNL_Tongues_m} in comparison to that displayed in figure~\ref{fig:C8_Faraday_tongues_comparison}, is clarified by the asymptotic relation~\eqref{eq:C8_asym_SH_m} and, particularly by the term $\left(\frac{\Omega+\sigma_{0,i}}{2\omega_0}-1\right)$. In \S\ref{sec:C8_Sec1}, the analysis is based on a fixed forcing frequency, while the wavenumber $k$ and, hence, the natural frequency $\omega_0$, are let free to vary. The first sub-harmonic Faraday tongue occurs when $\Omega+\sigma_{0,i} \approx 2\omega_0$. Since $\Omega$ is fixed and $\sigma_{0,i}>0$, $\Omega+\sigma_{0,i}>\Omega$ such that $\omega_0$ and therefore $k$ have to increase in order to satisfy the relation. On the other hand, if the wavenumber $m$ and, hence, $\omega_0$ are fixed as in this section, then $2\omega_0-\sigma_{0,i}<2\omega_0$ and the forcing frequency around which the sub-harmonic resonance is centered, decreases of a contribution $\sigma_{0,i}$, which introduces a frequency detuning responsible for the negative frequency shift displayed in figure~\ref{fig:C8_Fig_WNL_Tongues_m}.\\

\subsection{Discussion on the system's spatial quantization}\label{subsec:C8_Sec2sub2}

A first aspect that needs to be better discussed is the frequency-dependence of the damping coefficient $\sigma_{n}$ associated with each Faraday's tongue. In the case of horizontally infinite cells, the most natural description for investigating the system's stability properties is in the $\left(k,f\right)$ plane for a fixed forcing angular frequency $\Omega$ \citep{kumar1994parametric}. According to our model, the oscillating system's response occurring within each tongue is characterized by a Stokes boundary layer thickness $\delta_n=\sqrt{2\nu/\left(n\Omega+\alpha\right)}/b$. For instance, let us consider sub-harmonic resonances with $\alpha=\Omega/2$. As $\Omega$ is fixed (see any sub-panel of figure~\ref{fig:C8_Faraday_tongues_comparison}), each unstable region sees a constant $\delta_n$ (with $n=0,1,2,\hdots$) and hence a constant damping $\sigma_n$.\\
\indent  On the other hand, in the case of quantized wavenumber as for the annular cell of \S\ref{sec:C8_Sec2}, the most suitable description is in the driving frequency-driving amplitude plane at fixed wavenumber $m$ (see figure~\ref{fig:C8_Fig_WNL_Tongues_m}) \citep{batson2013faraday}. In this description, each sub-harmonic ($\alpha=\Omega/2$) or harmonic ($\alpha=\Omega$) $n$th tongue associated with a wavenumber $m$, sees a $\delta_n$, and thus a $\sigma_n$, changing with $\Omega$ along the tongue itself.\\
\indent  Furthermore, it is important to realize that in a real lab-scale experiment, the horizontal size of rectangular cells is never actually infinite. It follows that if the analysis of \S\ref{sec:C8_Sec1} is restrained to horizontally finite cells of overall length $l$, then one must impose the no-penetration condition for $u'$ at $x'=\pm l/2$, which would set the admissible wavenumbers to $k=m\pi/l$ only, with $m=0,1,2,\hdots\in\mathbb{N}$, thus completing the analogy with the annular configuration.\\
\indent  In such a case however, the solution form~\eqref{eq:C8_Darcy_like_eq} prevents the no-slip condition for the in-plane $xz$-velocity components to be imposed \citep{viola2017sloshing}. This always translates into an underestimation of the overall damping of the system in standard Hele-Shaw cells, although the sidewall contribution is expected to be negligible for sufficiently long cells.\\
\indent  On the other hand, the case of a thin annulus, by naturally filtering out this extra dissipation owing to the periodicity condition, offers a prototype configuration that can allow one to better quantify the correction introduced by the present gap-averaged model when compared to dedicated experiments, as outlined in the next section.

\section{Experiments}\label{sec:C8_EXP}

\subsection{Setup}\label{subsec:C8_EXP1}

\begin{figure}
\centering
\includegraphics[width=1\textwidth]{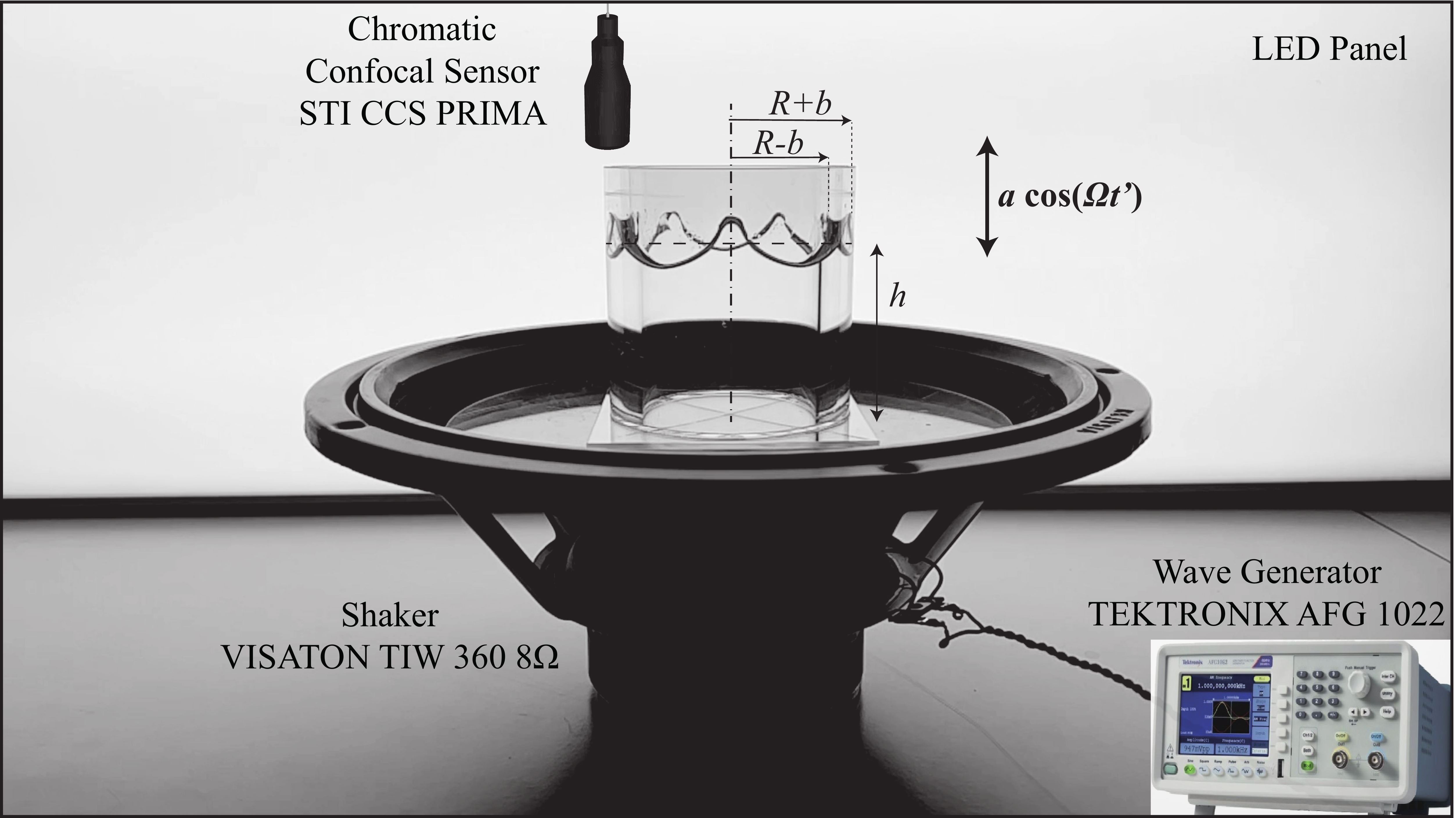}
\caption{Photo of the experimental setup}
\label{fig:C8_Exp_Setup} 
\end{figure}

The experimental apparatus, shown in figure~\ref{fig:C8_Exp_Setup}, is very simple. We used a Plexiglas annular container of height $100\,\text{mm}$, nominal radius $R=44\,\text{mm}$ and gap-size $b=7\,\text{mm}$, which is then filled to a depth $h=65\,\text{mm}$ with ethanol 70\% (see table~\ref{tab:FluidProp} for the fluid properties). An air conditioning system helps in maintaining the temperature of the room at around $22^{\circ}$. The container is mounted on a loudspeaker VISATON TIW 360 8$\Omega$ placed on a flat table and connected to a wave generator TEKTRONIX AFG 1022, whose output signal is amplified using a wideband amplifier THURKBY THANDER WA301. The motion of the free surface is recorded with a digital camera NIKON D850 coupled with a 60mm f/2.8D lens and operated in slow motion mode, allowing for an acquisition frequency of 120 frames per second. A LED panel placed behind the apparatus provides back illumination of the fluid interface for better optimal contrast. The wave generator imposes a sinusoidal alternating voltage, $v=\left(Vpp/2\right)\cos{\left(\Omega t'\right)}$, with $\Omega$ the angular frequency and $Vpp$ the full peak-to-peak voltage. The response of the loudspeaker to this input translates into a vertical harmonic motion of the container, $a\cos{\left(\Omega t'\right)}$, whose amplitude, $a\,\left[\text{mm}\right]$, is measured with a chromatic confocal displacement sensor STI CCS PRIMA/CLS-MG20. This optical pen, which is placed around $2\,\text{cm}$ (within the admissible working range of $2.5\,\text{cm}$) above the container and points at the top flat surface of the outer container's wall, can detect the time-varying distance between the fixed sensor and the oscillating container's surface with a sampling rate in the order of kHz and a precision of $\pm1\,\mu\text{m}$. Therefore, the pen can be used to obtain a very precise real-time value of $a$ as the voltage amplitude $Vpp$ and the frequency $\Omega$ are adjusted.

\subsection{Identification of the accessible experimental range}\label{subsec:C8_EXP11}

\begin{figure}
\centering
\includegraphics[width=1\textwidth]{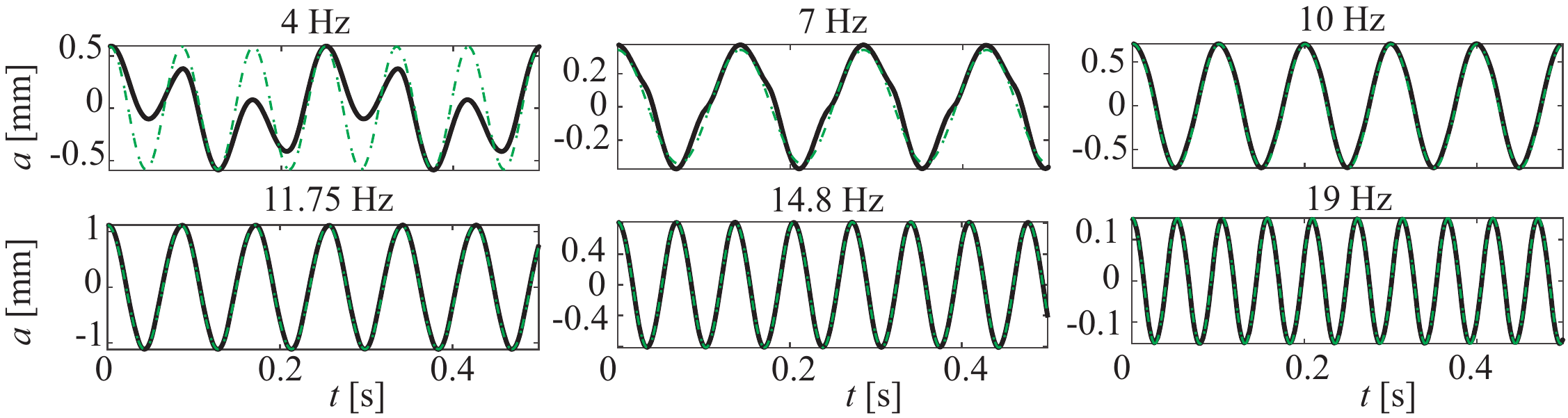}
\includegraphics[width=1\textwidth]{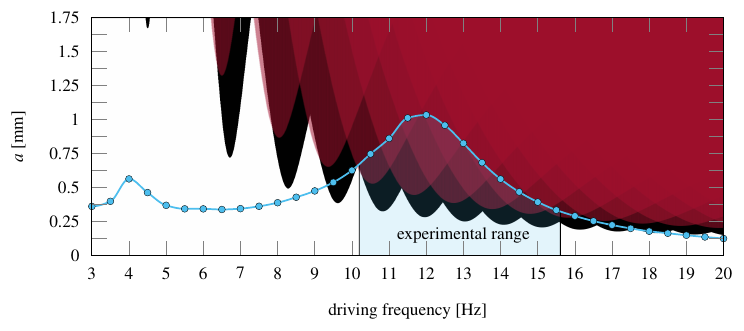}
\caption{\textit{Top}: vertical container displacement $a$ versus time at different forcing frequencies. The black curves are the measured signal, while the green dash-dotted curves are sinusoidal fitting. Below a forcing frequency of 8 Hz, the loudspeaker's output begins to depart from a sinusoidal signal. \textit{Bottom}: same as in figure~\ref{fig:C8_Fig_WNL_Tongues_m}(d): sub-harmonic Faraday tongues computed by accounting for contact line dissipation with a mobility parameter $M=0.0485$. The light blue curve here superposed corresponds to the maximal vertical displacement $a$ achievable with our setup. With this constraint, Faraday waves are expected to be observable only in the frequency range highlighted in blue.}
\label{fig:C8_ExpRangeCheckSin} 
\end{figure}

Our simple setup put some constraints on the explorable experimental frequency range.\\
\indent  (i) First, we need to ensure that the loudspeaker's output translates into a vertical container's displacement following a sinusoidal time signal. To this end, the optical sensor is used to measure the container motion at different driving frequencies. These time signals are then fitted with a sinusoidal law. Figure~\ref{fig:C8_ExpRangeCheckSin} shows how below a forcing frequency of 8 Hz, the loudspeaker's output begins to depart from a sinusoidal signal. This check imposes a first lower bound on the explorable frequency range.\\
\indent  (ii) In addition, as Faraday waves only appear above a threshold amplitude, it is convenient to measure \textit{a priori} the maximal vertical displacement $a$ achievable. The loudspeaker response curve is reported in the bottom part of figure~\ref{fig:C8_ExpRangeCheckSin}. A superposition of this curve with the predicted Faraday's tongues immediately identifies the experimental frequency range within which the maximal achievable $a$ is larger than the predicted Faraday threshold so that standing waves are expected to emerge in our experiments. Assuming the herein proposed gap-averaged model (red regions) to give a good prediction of the actual instability onset, the experimental range explored in the next section is limited to approximately $\in\left[10.2,15.6\right]$ Hz.

\subsection{Procedure}\label{subsec:C8_EXP12}

\begin{figure}
\centering
\includegraphics[width=1\textwidth]{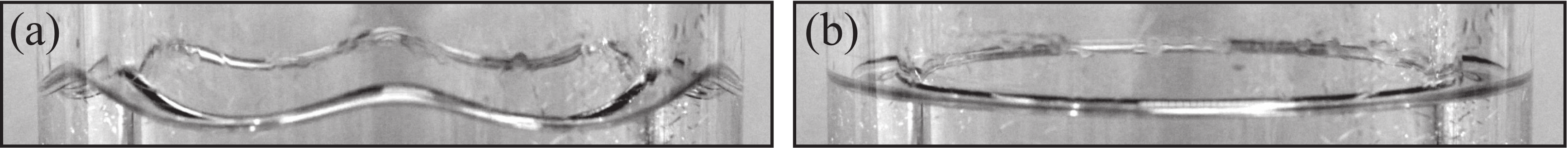}
\caption{Free surface shape at a forcing frequency $1/T=11.7\,\text{Hz}$ and corresponding to: (a) the lowest forcing amplitude value, $a=0.4693\,\text{mm}$, for which the $m=6$ standing wave is present (the figure shows a temporal snapshot); (b) the largest forcing amplitude value, $a=0.4158\,\text{mm}$, for which the surface becomes flat and stable again. Despite the small forcing amplitude variation, the change in amplitude is large enough to allow for a visual inspection of the instability threshold with sufficient accuracy.}
\label{fig:C8_Visual_Inspection} 
\end{figure}

Given the constraints discussed in \S\ref{subsec:C8_EXP11}, experiments have been carried out in a frequency range between 10.2 Hz and 15.6 Hz with a frequency step of 0.1 Hz. For each fixed forcing frequency, the Faraday threshold is determined as follows: the forcing amplitude $a$ is set to the maximal value achievable by the loudspeaker, so as to quickly trigger the emergence of the unstable Faraday wave. The amplitude is then progressively decreased until the wave disappears and the surface becomes flat again.\\
\indent  More precisely, a first quick pass across the threshold is made to determine an estimate of the sought amplitude. A second pass is then made by starting again from the maximum amplitude and decreasing it. When we approach the value determined during the first pass, we perform finer amplitude decrements, and we wait several minutes between each amplitude change to ensure that the wave stably persists. We eventually identify two values: the last amplitude where the instabilities were present (see figure~\ref{fig:C8_Visual_Inspection}(a)) and the first one where the surface becomes flat again (see figure~\ref{fig:C8_Visual_Inspection}(b)). Two more runs following an identical procedure are then performed to verify the values previously found. Lastly, an average between the smallest unstable amplitude and the largest stable one gives us the desired threshold.\\
\indent  Once the threshold amplitude value is found for the considered frequency, the output of the wave generator is switched off, the frequency is changed, and the steps presented above are implemented again for the new frequency. In this way we always start from a stable configuration, hence limiting the possibility of nonlinear interaction between different modes.\\
\indent For each forcing frequency, the two limiting amplitude values, identified as described above, are used to define the error bars reported in figure~\ref{fig:C8_comparison_exp}. Those error bars must also account for the optical pen’s measurement error ($0.1\,\text{$\mu$m}$), as well as the non-uniformity of the output signal. By looking at the measured average, minimum, and maximum amplitude values in the temporal output signal, it is noteworthy that the average value typically deviates from the minimum and maximum by around $10\,\text{$\mu$m}$. Consequently, we incorporate in the error bars this additional $10\,\text{$\mu$m}$ of uncertainty in the value of $a$. The uncertainty in the frequency of the output signal is not included in the definition of the error bars, as it is extremely small, on the order of 0.001 Hz.\\

\subsection{Instability onset and wave patterns}\label{subsec:C8_EXP2}

\begin{figure}
\centering
\includegraphics[width=1\textwidth]{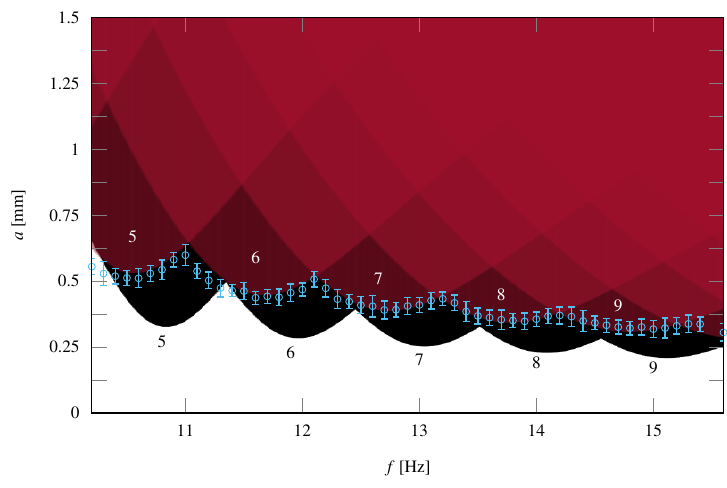}
\caption{Experiments (empty circles) are compared to the theoretically predicted sub-harmonic Faraday threshold computed via Floquet analysis~\eqref{eq:C8_Mathieu1_eig} for different fixed azimuthal wavenumber $m$ and according to the standard (black region) and revised (red regions) gap-averaged models. The shaded band around the instability onset indicates the error bar for the threshold amplitudes at each measured driving frequency. The tongues are computed by including contact line dissipation with a value of $M$ equal to $0.0485$ as in figures~\ref{fig:C8_Fig_WNL_Tongues_m}(b)-(d) and~\ref{fig:C8_ExpRangeCheckSin}. As explained in \S\ref{subsec:C8_EXP12}, the vertical error bars indicate the amplitude range between the smallest measured forcing amplitude at which the instability was detected and the largest one at which the surface remains stable and flat. These two limiting values are successively corrected by accounting for the ptical pen’s measurement error and the non-uniformity of the output signal of the loudspeaker.}
\label{fig:C8_comparison_exp} 
\end{figure}

\begin{figure}
\centering
\includegraphics[width=1\textwidth]{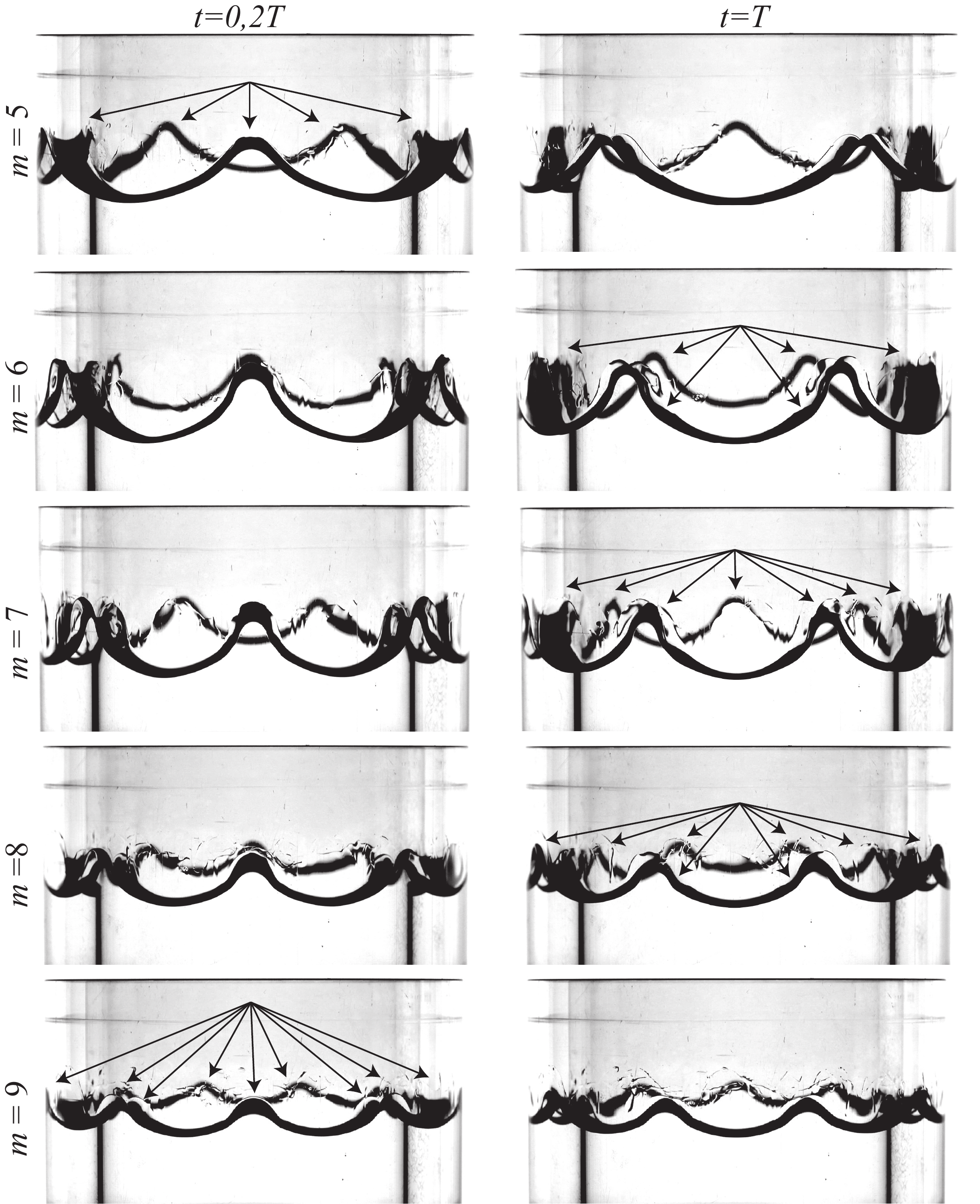}
\caption{Snapshots of the wave patterns experimentally observed within the sub-harmonic Faraday tongues associated with the azimuthal wavenumbers $m=5,6,7,8$ and $9$. $T$ is the forcing period, which is approximately half the oscillation period of the wave response. These patterns appear for: ($m=5$) $1/T=10.6\,\text{Hz}$, $a=0.8\,\text{mm}$; ($m=6$) $1/T=11.6\,\text{Hz}$, $a=1.1\,\text{mm}$; ($m=7$) $1/T=12.7\,\text{Hz}$, $a=0.9\,\text{mm}$; ($m=8$), $1/T=13.7\,\text{Hz}$, $a=0.6\,\text{mm}$; ($m=9$) $1/T=14.8\,\text{Hz}$, $a=0.4\,\text{mm}$. These forcing amplitudes are the maximal achievable at their corresponding frequencies (see figure~\ref{fig:C8_ExpRangeCheckSin} for the associated operating points). The number of peaks is easily countable by visual inspection of two time snapshots of the oscillating pattern exactred at $t=0,T$ and $t=T/2$. This provides a simple criterion for the identification of the resonant wavenumber $m$. See also supplementary movies 1-5 at: LINK.}
\label{fig:C8_wave_m56789} 
\end{figure}

The experimentally detected threshold at each measured frequency is reported in figure~\ref{fig:C8_comparison_exp} in terms of forcing acceleration $f$ and amplitude $a$. Once again, the black unstable regions are calculated according to the standard gap-averaged model with $\sigma_{BL}=12\nu/b^2$, whereas red regions are the unstable tongues computed using the modified damping $\sigma_{BL}=\chi_n\nu/b^2$. Both scenarios include contact line dissipation $\sigma_{CL}=\left(2M/\rho b\right)\left(m/R\right)\tanh{\left(mh/R\right)}$, with a value of $M$ equal to $0.0485$ for ethanol 70\%. Although, at first, this value has been simply selected in order to fit well our experimental measurements, it is in perfect agreement with the linear relation linking $M$ to the liquid's surface tension reported in figure~5 of \cite{hamraoui2000can} and used by \cite{li2019stability} (see table~\ref{tab:FluidProp}).\\
\indent  As figure~\ref{fig:C8_comparison_exp} strikingly shows, the present theoretical thresholds match well our experimental measurements. On the contrary, the poor description of the oscillating boundary layer in the classical Darcy model translates into a lack of dissipation. The arbitrary choice of a higher fitting parameter $M$ value, e.g. $M\approx0.09$ would increase contact line dissipation and compensate for the underestimated Stokes boundary layer one, hence bringing these predictions much closer to experiments; however, such a value would lie well beyond the typical values reported in the literature. Furthermore, the real damping coefficient $\sigma_{BL}=12\nu/b^2$ given by the Darcy theory does not account for the frequency detuning displayed by experiments. This frequency shift is instead well captured by the imaginary part of the new damping $\sigma_{BL}=\chi_n\nu/b^2$ ($\chi_n=\chi_{n,r}+\text{i}\chi_{n,i}$).\\
\indent  Within the experimental frequency range considered, five different standing waves, corresponding to $m=5,6,7,8$ and $9$, have emerged. The identification of the wavenumber $m$ has been simply performed by visual inspection of the free surface patterns reported in figure~\ref{fig:C8_wave_m56789}. Indeed, by looking at two time snapshots separated by a forcing period $T$, it is possible to count the various wave peaks along the azimuthal direction.\\ 
\indent When looking at figure~\ref{fig:C8_comparison_exp}, it is worth commenting that on the left sides of the marginal stability boundaries associated with modes $m=5$ and $6$ we still have a little discrepancy between experiments and the model. Particularly, the experimental thresholds are slightly lower than the predicted ones. A possible explanation can be given by noticing that our experimental protocol is agnostic to the possibility of subcritical bifurcations and hysteresis, while such behaviour has been predicted by \cite{Douady90}.\\
\indent  As a last comment, one has to keep in mind that the Hele-Shaw approximation remains good only if the wavelength, $m/2\pi R$ does not become too small, i.e. comparable to the cell's gap, $b$. In other words, one must check that the ratio $m b/2\pi R$ is of the order of the small separation-of-scale parameter, $\epsilon$. For the largest wavenumber observed in our experiments, $m=9$, the ratio $m b/2\pi R$ amounts to 0.23, which is not exactly small. Yet, the Hele-Shaw approximation is seen to remain fairly good.

\subsection{Contact angle variation and thin film deposition}\label{subsec:C8_EXP3}

\begin{figure}
\centering
\includegraphics[width=1\textwidth]{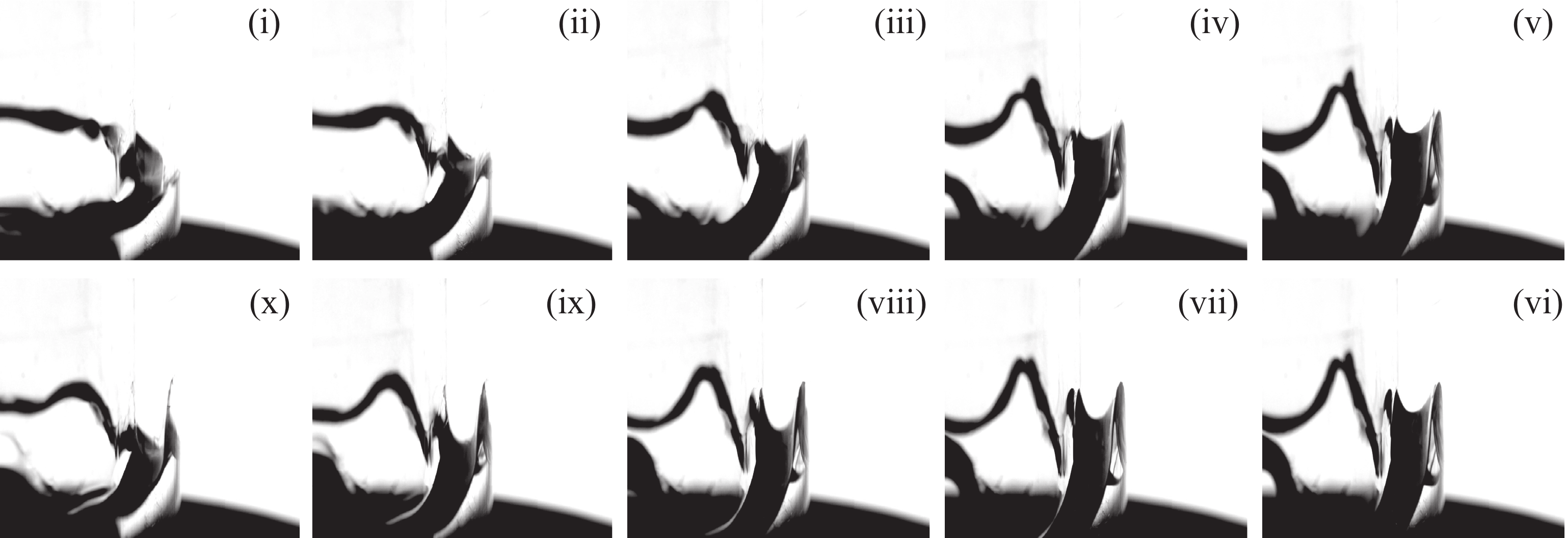}
\caption{Zoom of the meniscus dynamics recorded at a driving frequency $11.6\,\text{Hz}$ and amplitude $a=1.2\,\text{mm}$ for $m=6$. Seven snapshots, (i)-(vii), covering one oscillation period, $T$, for the container motion are illustrated. These snapshots show how the meniscus profile and the macroscopic contact angle change in time during the second half of the advancing cycle and the first half of the receding cycle, hence highlighting the importance of the out-of-plane curvature or capillary effects. . See also supplementary movie 6 at: LINK.}
\label{fig:C8_AngleDyn} 
\end{figure}

\begin{figure}
\centering
\includegraphics[width=1\textwidth]{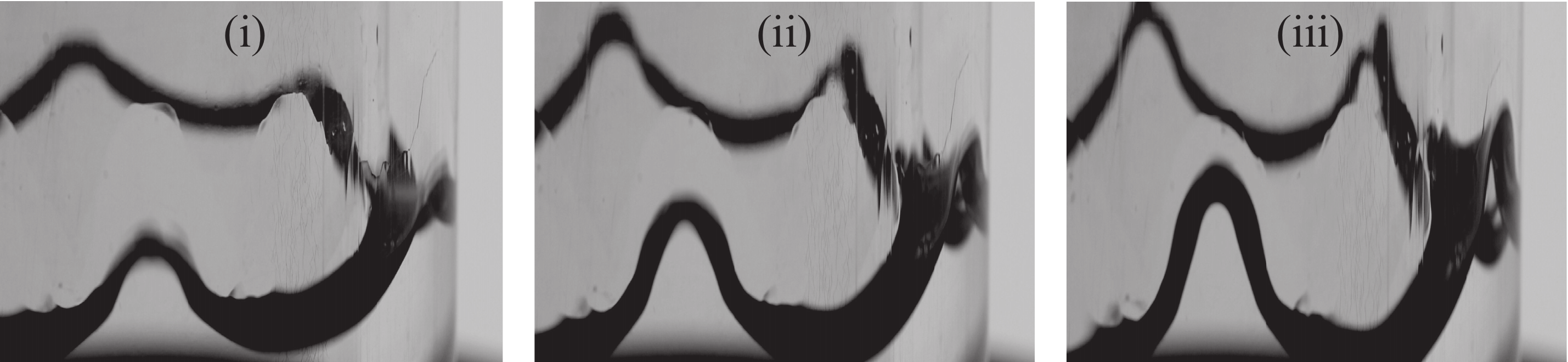} 
\caption{These three snapshots correspond to snapshots (ii), (iii) and (iv) of figure~\ref{fig:C8_AngleDyn} and show, using a different light contrast, how the contact line constantly moves over a wetted substrate due to the presence of a stable thin film deposited and alimented at each cycle. See also supplementary movies 7 at: LINK.}
\label{fig:C8_film} 
\end{figure}

Before concluding, it is worth commenting on why the use of dynamic contact angle model~\eqref{eq:C8_contact_line_mod} is justifiable and seen to give good estimates of the Faraday thresholds.\\
\indent  Existing lab experiments have revealed that liquid oscillations in Hele-Shaw cells constantly experience an up-and-down driving force with an apparent contact angle $\theta$ constantly changing \citep{jiang2004contact}. Our experiments are consistent with such evidence. In figure~\ref{fig:C8_AngleDyn} we report seven snapshots, (i)-(vii), covering one oscillation period, $T$, for the container motion. These snapshots illustrate a zoom of the dynamic meniscus profile and show how the macroscopic contact angle changes in time during the second half of the advancing cycle (i)-(v) and the first half of the receding cycle (vi)-(x), hence highlighting the importance of the out-of-plane meniscus curvature variations. Thus, on the basis of our observations, it seems appropriate to introduce in the theory a contact angle model so as to justify this associated additional dissipation, which would be neglected by assuming $M=0$. The model used in this study, and already implemented by \cite{li2019stability}, is very simple; it assumes the cosine of the dynamic contact angle to linearly depend on the contact line speed through the capillary number $Ca$ \citep{hamraoui2000can}. Accounting for such a model is shown, both in \cite{li2019stability} and in this study, to supplement the theoretical predictions by a sufficient extra dissipation suitable to match experimental measurements.\\
\indent  This dissipation eventually reduces to a simple damping coefficient $\sigma_{CL}$ as it is of linear nature. A unique constant value of the mobility parameter $M$ is sufficient to fit all our experimental measurements at once, suggesting that the meniscus dynamics is not significantly affected by the evolution of the wave in the azimuthal direction, i.e. by the wavenumber, and $M$ can be seen as an intrinsic property of the liquid-substrate interface.\\
\indent  Several studies have discussed the dependence of the system's dissipation on the substrate material \citep{Huh71,Dussan79,Cocciaro93,ting1995boundary,Eral2013,Viola2018a,Viola2018b,xia2018moving}. These authors, among others, have unveiled and rationalized interesting features such as solid-like friction induced by contact angle hysteresis. This strongly nonlinear contact line behaviour does not seem to be present in our experiments. This can be tentatively explained by looking at figure~\ref{fig:C8_film}. These snapshots illustrate how the contact line constantly flows over a wetted substrate, due to the presence of a stable thin film deposited and alimented at each oscillation cycle. This feature has been also recently described by \cite{dollet2020transition}, who showed that the relaxation dynamics of liquid oscillation in a U-shaped tube filled with ethanol, due to the presence of a similar thin film, obey an exponential law that can be well-fitted by introducing a simple linear damping, as done in this work.


\section{Conclusions}\label{sec:C8_CONC}

Previous theoretical analyses for Faraday waves in Hele-Shaw cells have so far relied on the Darcy approximation, which is based on the parabolic flow profile assumption in the narrow direction and that translates into a real-valued damping coefficient $\sigma_{BL}=12\nu/b^2$, with $\nu$ the fluid kinematic viscosity and $b$ the cell's gap-size, that englobes the dissipation originated from the Stokes boundary layers over the two lateral walls. However, Darcy’s model is known to be inaccurate whenever inertia is not negligible, e.g. in unsteady flows such as oscillating standing or traveling  waves.\\
\indent  In this work, we have proposed a gap-averaged linear model that accounts for inertial effects induced by the unsteady terms in the Navier-Stokes equations, amounting to a pulsatile flow where the fluid’s motion reduces to a two-dimensional oscillating, reminiscent of the Womersley flow in cylindrical pipes. When gap-averaging the linearized Navier-Stokes equation, this results in a modified damping coefficient, $\sigma_{BL}=\chi_n\nu/b^2$, with $\chi_n=\chi_{n,r}+i\chi_{n,i}$ complex-valued, which is a function of the ratio between the Stokes boundary layer thickness and the cell’s gap-size, and whose value depends on the frequency of the system’s response specific to each unstable parametric Faraday tongue.\\
\indent  After having revisited the ideal case of infinitely long rectangular Hele-Shaw cells, we have considered the case of Faraday waves in thin annuli. This annular geometry, owing to the periodicity condition, naturally filters out the additional, although small, dissipation coming from the lateral wall in the elongated direction of finite-size lab-scale Hele-Shaw cells. Hence, a thin annulus offers a prototype configuration that can allow one to better quantify the correction introduced by the present gap-averaged theory when compared to dedicated experiments and to the standard gap-averaged Darcy model.\\
\indent  A series of homemade experiments for the latter configuration has proven that Darcy's model typically underestimates the Faraday threshold, as $\chi_{n,r}>12$, and overlooks a frequency detuning introduced by $\chi_{n,i}>0$, which appears essential to correctly predict the location of the Faraday's tongue in the frequency spectrum. The frequency-dependent gap-averaged model here proposed successfully predicts these features and brings the Faraday thresholds estimated theoretically closer to the ones measured.\\
\indent  Furthermore, a close look at the experimentally observed meniscus and contact angle dynamics clearly highlighted the importance of the out-of-plane curvature, whose contribution has been neglected so far in the literature, with the exception of \cite{li2019stability}. This evidence justifies the employment of a dynamical contact angle model to recover the extra contact line dissipation and close the gap with experimental measurements.\\
\indent A natural extension of this work is to examine the existence of a drift instability at higher forcing amplitudes. 


\subsubsection*{\textbf{\textup{Supplementary Material}}}
\indent Supplementary movies 1-5 show the time evolution of the free surface associated with the snapshots reported in figure~\ref{fig:C8_wave_m56789}. Supplementary movies 6 and 7 provide instead better visualizations of the meniscus and the thin film, dynamics, respectively, as illustrated in figures~\ref{fig:C8_AngleDyn} and~\ref{fig:C8_film} of this manuscript. Supplementary movies are available at link: LINK.

\subsubsection*{\textbf{\textup{Funding}}}
\indent We acknowledge the Swiss National Science Foundation under grant 178971.

\subsubsection*{\textbf{\textup{Declaration of Interests}}}
\indent The authors report no conflict of interest.

\subsubsection*{\textbf{\textup{Author Contributions}}}
\indent A. B., F. V. and F. G. created the research plan. A.B. formulated analytical and numerical models. A.B. led model solution. A. B. and B. J. designed the experimental setup. B. J. performed all experiments. A.B., B. J., F.V. and F.G. wrote the manuscript.


\bibliographystyle{jfm}
\bibliography{Bibliography}

\end{document}